%% file: main.tex
\documentclass[conference]{IEEEtran}
\usepackage{fancyhdr}
\usepackage[normalem]{ulem}
\usepackage[hyphens]{url}

\usepackage{graphicx} 
\usepackage{subfigure} 

\usepackage[numbers]{natbib}

\usepackage{algorithm}
\usepackage{algorithmic}

\usepackage{xspace}
\usepackage{subfigure}
\usepackage{amsmath}
\usepackage{mathtools}

\usepackage[font=small,labelfont=bf]{caption}

\usepackage{color}

\newcommand{\old}[1]{}

\newcommand{\vdnn}[0]{\texttt{vDNN}\xspace}

\newcommand{\fig}[1]{Figure~\ref{#1}}
\newcommand{\sect}[1]{Section~\ref{#1}}

\newcommand{\eqn}[1]{Equation~\ref{#1}}

\newcommand{\featureIn}[0]{\texttt{X}\xspace}
\newcommand{\featureOut}[0]{\texttt{Y}\xspace}
\newcommand{\gradientIn}[0]{\texttt{dY}\xspace}
\newcommand{\gradientOut}[0]{\texttt{dX}\xspace}
\newcommand{\featureIns}[0]{\texttt{X}s\xspace}

\newcommand{\gradientIns}[0]{\texttt{dY}s\xspace}

\newcommand{\weight}[0]{\texttt{W}\xspace}
\newcommand{\workspace}[0]{\texttt{WS}\xspace}

\newcommand{\streamMemory}[0]{\texttt{stream$_{memory}$}\xspace}
\newcommand{\streamCompute}[0]{\texttt{stream$_{compute}$}\xspace}

\newcommand{\vdnnAll}[0]{\texttt{vDNN$_{all}$}\xspace}
\newcommand{\vdnnAllM}[0]{\texttt{vDNN$_{all}(m)$}\xspace}

\newcommand{\vdnnConv}[0]{\texttt{vDNN$_{conv}$}\xspace}
\newcommand{\vdnnDyn}[0]{\texttt{vDNN$_{dyn}$}\xspace}

\newcommand{\layer}[1]{layer$_{(#1)}$\xspace}

\newcommand\blfootnote[1]{%
\begingroup
\renewcommand\thefootnote{}\footnote{#1}%
\addtocounter{footnote}{-1}%
\endgroup
}

\ifCLASSINFOpdf
\else
\fi
\renewcommand{\baselinestretch}{.999}
\begin{document}

%
\title{vDNN: Virtualized Deep Neural Networks for \\Scalable, Memory-Efficient Neural Network Design}

\author{
\IEEEauthorblockN{
Minsoo Rhu\hspace{2em}Natalia Gimelshein\hspace{2em}Jason Clemons\hspace{2em}Arslan Zulfiqar\hspace{2em}Stephen W. Keckler}
\IEEEauthorblockA{
NVIDIA\\
Santa Clara, CA 95050\\
\texttt{\{mrhu, ngimelshein, jclemons, azulfiqar, skeckler\}@nvidia.com}\\
}
}

%



\maketitle

\input{tex/abstract}


%
\IEEEpeerreviewmaketitle

\blfootnote{
Published as a conference paper at the $49^{\text{th}}$ IEEE/ACM International Symposium on Microarchitecture (MICRO-49), 2016.
}

\input{tex/intro}

\input{tex/background}

\input{tex/vdnn}

\input{tex/eval}

\input{tex/results}

\input{tex/related}

\input{tex/conclusion}


\section*{Acknowledgment}
We thank our colleagues at NVIDIA for their
feedback on this work, and in particular John Tran, Sharan Chetlur, Simon Layton, 
and Cliff Woolley for their contributions to \vdnn concepts and infrastructure.



%

\bibliographystyle{ieeetr}
\bibliography{main}




\end{document}

%% file: tex/abstract.tex
\begin{abstract} 

The most widely used machine learning frameworks require users to
carefully tune their memory usage so that the deep neural network (DNN)
fits into the DRAM capacity of a GPU. This restriction hampers
a researcher's flexibility to study different machine learning algorithms,
forcing them to either use a less desirable network architecture
or parallelize the processing across multiple GPUs.
We propose a runtime memory manager that virtualizes the memory usage of DNNs
such that both GPU and CPU memory can simultaneously be utilized for training
larger DNNs. Our virtualized DNN (\vdnn) reduces the average GPU memory
usage of AlexNet by up to 89\%, OverFeat by 91\%, 
and GoogLeNet by 95\%, a significant reduction in
memory requirements of DNNs. Similar experiments on VGG-16, one of the deepest
and memory hungry DNNs to date, demonstrate the memory-efficiency
of our proposal. \vdnn enables VGG-16 with batch size 256 (requiring 28 GB
of memory) to be trained on a single NVIDIA Titan X GPU card containing 12 GB
of memory, with 18\% performance loss compared to a hypothetical, oracular GPU with
enough memory to hold the entire DNN.

\end{abstract}

%% file: tex/intro.tex
\section{Introduction}

Deep neural networks (DNNs) have recently been successfully deployed in various
application domains such as computer vision~\cite{alexnet}, speech
recognition~\cite{graves:2005:fpc}, and natural language
processing~\cite{collobert:2011:nlp_from_scratch} thanks to their superior
performance compared to traditional state-of-the-art approaches.  Such
proliferation of deep learning techniques has led several software
frameworks to be developed in recent years to analyze and facilitate the design
of neural networks~\cite{tensorflow,torch,theano,caffe}. The list of available frameworks
continue to expand with developers constantly adding more features and
improving computational efficiency to foster research in the area of deep
learning. Due to the tremendous compute horsepower offered by graphics
processing units (GPUs), these frameworks provide strong backend support for
GPU software libraries such as cuDNN~\cite{cudnn_v4}. In
fact, almost every group today involved in training neural networks is
deploying GPUs for accelerated deep learning~\cite{bahrampour:2016:arxiv}.

\begin{figure}[t!] \centering
\includegraphics[width=0.48\textwidth]{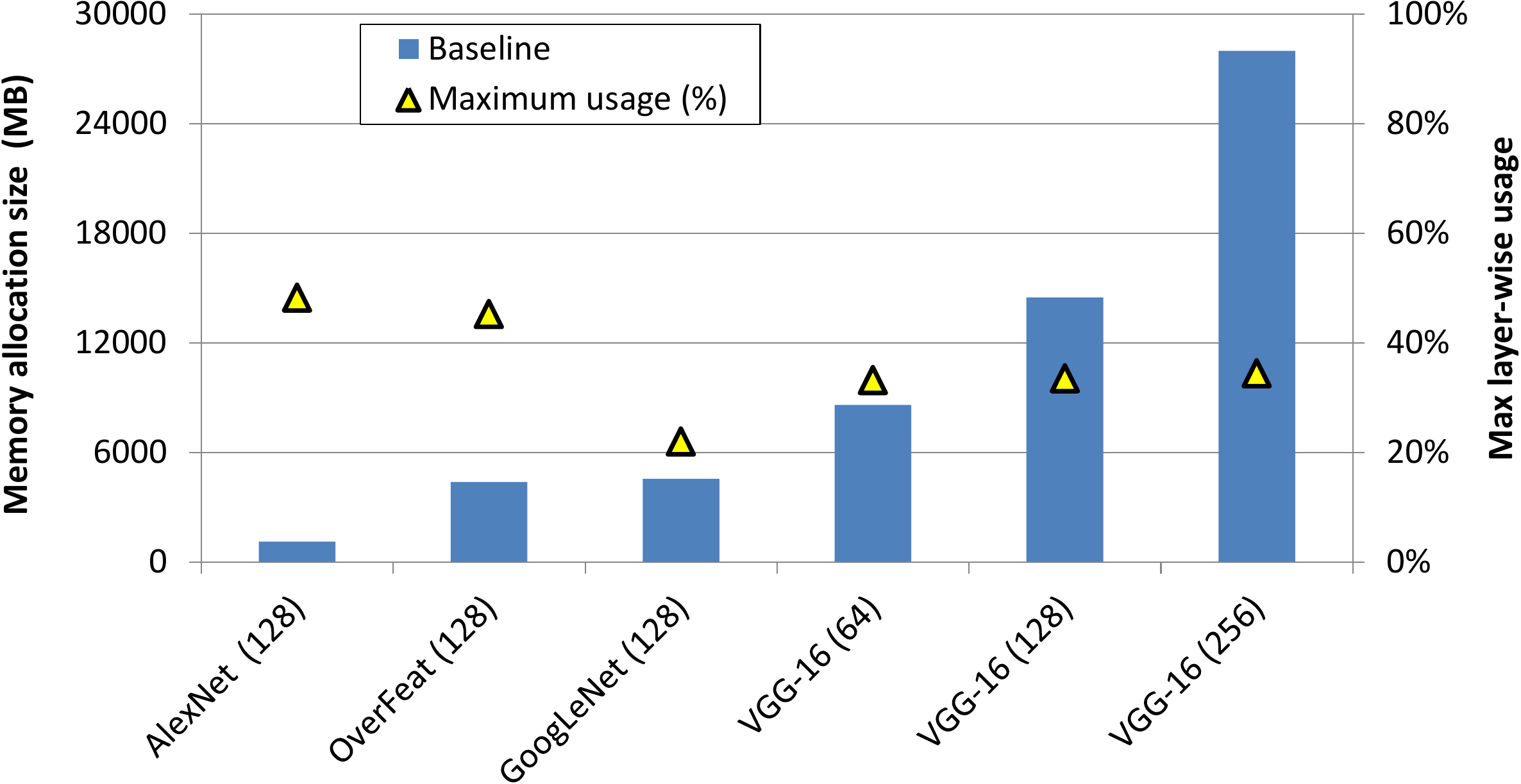}
\caption{GPU memory usage when using the baseline, network-wide allocation policy (left axis).
The right axis shows the maximum fraction of this baseline allocation actually utilized when 
traversing through the network layer-wise. The numbers next to the names of each network refer to the batch size
throughout this paper. Studied DNNs are detailed in \sect{sect:benchmark}.
}
\vspace{-1em}
\label{fig:mem_usage_global_vs_local}
\end{figure}

While these popular machine learning (ML) frameworks facilitate the study of
DNNs, a major limitation of the use of these frameworks is that the DRAM
capacity limits of the GPU(s) in the system eventually limit the size the of
the DNN that can be trained (\sect{sect:motivation}).  To work around the memory capacity
bottleneck~\cite{nextplatform_baidu,persistent_rnn}, ML practitioners must either use less desirable DNN
architectures (e.g., smaller number of layers, smaller batch sizes, less
performant but more memory-efficient convolutional algorithms) or parallelize
the DNN across multiple GPUs~\cite{alex_weird_trick}.
\fig{fig:mem_usage_global_vs_local} highlights how the memory consumption
trends of the ImageNet~\cite{imagenet} winning DNNs have evolved over time.
AlexNet~\cite{alexnet}, for instance, only contained 5 convolutional layers
with 2 fully-connected layers and required a ``mere'' $1.1$ GB of memory
allocation for training, which is well below the $12$ GB memory capacity of the
state-of-the-art NVIDIA Titan X.  The more recent VGG-16~\cite{vggnet}, on the
other hand, contains 16 convolutional layers and 3 fully-connected layers,
incurring a total of $28$ GB of memory usage for batch size 256. Because a
single GPU can only accommodate a batch size of 64 for VGG-16, training with
batch 256 requires parallelization across multiple GPUs or the network must
be sequentially executed multiple times with smaller batches.  With the most recent ImageNet
winning network adopting more than a hundred convolutional
layers~\cite{msr_2015}, the trend in deep learning is to move towards
larger and deeper network designs~\cite{vggnet,wired:msr_deeper_nn,googlenet,nn_stochastic_depth}. 
As a result, alleviating the rigid
physical memory limitations of GPUs is becoming increasingly important.

In this paper, we propose \emph{virtualized Deep Neural Network} (\vdnn), a
runtime memory management solution that virtualizes the memory usage of deep
neural networks across both GPU and CPU memories. Our \vdnn allows ML
practitioners to deploy larger and deeper networks beyond the physical capacity
of available GPUs, enabling them to focus more on their algorithms while the
system architecture and runtime system transparently manage the allocation,
placement, movement, and release of their data. The motivation behind \vdnn is
based on the following three key observations: 1) DNNs trained via stochastic
gradient-descent (SGD) are designed and structured with multiple
layers~\cite{lecun_gd}; 2) the training of these neural networks involves a
series of \emph{layer-wise} computations, the order of which is statically
fixed and repeated for millions to billions of iterations throughout the entire
training process; and 3) even though the GPU can, at any given time, only
process a single layer's computation (due to the layer-wise computational
characteristics of SGD-based DNN training), popular ML frameworks adopt a
\emph{network-wide} memory allocation policy because DNN training requires the
intermediate feature maps of all the layers in the network to be backed up in
GPU memory for gradient updates (\sect{sect:motivation}).  In other words,
existing memory management schemes \emph{overprovision} the memory allocations
to accommodate the usage of the \emph{entire} network layers, even though the
GPU is only using a subset of this allocation for the layer-wise requirements.
We observe that such memory underutilization issue becomes more severe for
deeper networks, leading to 53\% to 79\% of allocated memory not being
used at all at any given time (\fig{fig:mem_usage_global_vs_local}).  The goal
of \vdnn is to \emph{conservatively} allocate GPU memory for the
\emph{immediate} usage of a given layer's computation so that the maximum and
average memory usage is drastically reduced, allowing researchers to train
larger networks. To achieve this goal, \vdnn exploits the data dependencies of
allocated data structures, particularly the intermediate feature maps that
account for the majority of memory usage (\sect{sect:motivation}), and either
releases or moves these intermediate data between GPU and CPU memory.
Specifically, \vdnn either 1) aggressively \emph{releases} these feature maps
from the GPU memory if no further reuse exists, or 2) \emph{offloads} (and
later \emph{prefetches}) to (from) CPU memory if further reuse does exist but
is not immediately required. By exploiting the inter-layer memory access and
reuse patterns of DNNs, our \vdnn memory manager  intelligently overlaps the
normal DNN computations with the offload/prefetch/release operations,
effectively virtualizing the memory usage of DNNs with little to no performance
loss. The operations of \vdnn are completely transparent to programmers and
enable them to train larger and deeper neural networks that consume memory well
beyond the limits of physical memory of GPUs today.  The key contributions of
our work are:

\begin{itemize} 

\item This work is the first to present a detailed, quantitative analysis on GPU-based DNN 
\emph{training}, as opposed to recent literature targeting energy-efficient accelerators for  
DNN \emph{inference}~\cite{diannao,dadiannao,eyeriss,song:2015:eie,eyeriss_isca,redeye,minerva,dnn_pim_reram,isacc,cnvlutin}.

\item 
To the best of our knowledge, our work is the first
that provides an in-depth characterization study on the memory access
characteristics of DNNs and their effect on the GPU memory system 
from an architectural perspective.

\item This work identifies the key limitations of current ML frameworks' memory
management policies as they require the network-wide memory usage of the target
DNN to monolithically fit within the physical capacity of the GPU.  
We demonstrate this by showing that existing frameworks fail in training $6$ out of the $10$ studied DNNs 
when their memory allocation size ($14$ GB to $67$ GB) exceeds the GPU memory budget ($12$ GB
in NVIDIA's Titan X).

\item We propose, implement, and evaluate a runtime memory manager called
\vdnn that virtualizes the memory usage of neural networks across
CPU and GPU memories. Our \vdnn solution reduces the average GPU memory usage of these
$6$ memory hungry networks by 73\% to 98\%, allowing them to be trained on a single Titan X
card. Compared to a hypothetical, oracular GPU containing enough memory to hold the entire
DNN, \vdnn incurs 1\% to 18\% performance overhead.

\end{itemize}

%% file: tex/background.tex
\section{Background and Motivation}
\label{sect:background}

This section provides an overview of modern DNNs, the memory management
policies of current ML frameworks, and their key limitations that
motivate this work.

\subsection{DNN Architecture}
\label{sect:dnn_arch}

Convolutional neural networks are one of the most popular ML algorithms for
high accuracy computer vision tasks. While other types of networks are also
gaining tractions (e.g., recurrent neural networks for natural language
processing), all of these DNNs are trained using a \emph{backward propagation}
algorithm~\cite{lecun_gd} via \emph{stochastic gradient-descent} (SGD).  For
clarity of exposition and owing to their state-of-the-art performance in the
ImageNet competition, this paper mainly focuses on the feedforward style
convolutional neural networks commonly seen in AlexNet~\cite{alexnet},
OverFeat~\cite{overfeat}, GoogLeNet~\cite{googlenet}, and VGG~\cite{vggnet}.
However, the key intuitions of our work are equally applicable to \emph{any}
neural network that exhibits \emph{layer-wise} computational characteristics
and is trained via SGD, detailed later in this section.

DNNs are designed using a combination of multiple
types of layers, which are broadly categorized as convolutional layers (CONV),
activation layers (ACTV), pooling layers (POOL), and fully-connected layers
(FC). A neural network is structured as a sequence of multiple instances of
these layers. DNNs for computer vision tasks in particular are
broadly structured into the following two modules: 1) the \emph{feature
extraction layers} that detect distinguishable features across input images,
and 2) the \emph{classification layers} that analyze the extracted features and
classify the image into a given image category. Feature extraction layers are
generally designed using CONV/ACTV/POOL layers and are positioned as the initial
part of the DNN. The classification layers are built up using the FC layers and
are found at the end of the DNN computation sequence. The general trend in deep
learning is to design the network with a large number of feature extraction
layers so that a deep hierarchy of features are trained for robust image
classification~\cite{vggnet,msr_2015,googlenet}. 

\vspace{1em}

\subsection{DNN Training vs. Inference}
\label{sect:dnn_training}

A neural network needs to be \emph{trained} before it can be deployed for an
\emph{inference} or \emph{classification} task. Training entails learning and
updating the \emph{weights} of the layers of a neural network by performing the
operations of forward and backward propagation
algorithms~\cite{lecun_gd}.  The direction of traversal, as well as the
mathematical operations that must be performed, differ for forward and
backward propagation. 

\begin{figure}[t!] \centering
\includegraphics[width=0.48\textwidth]{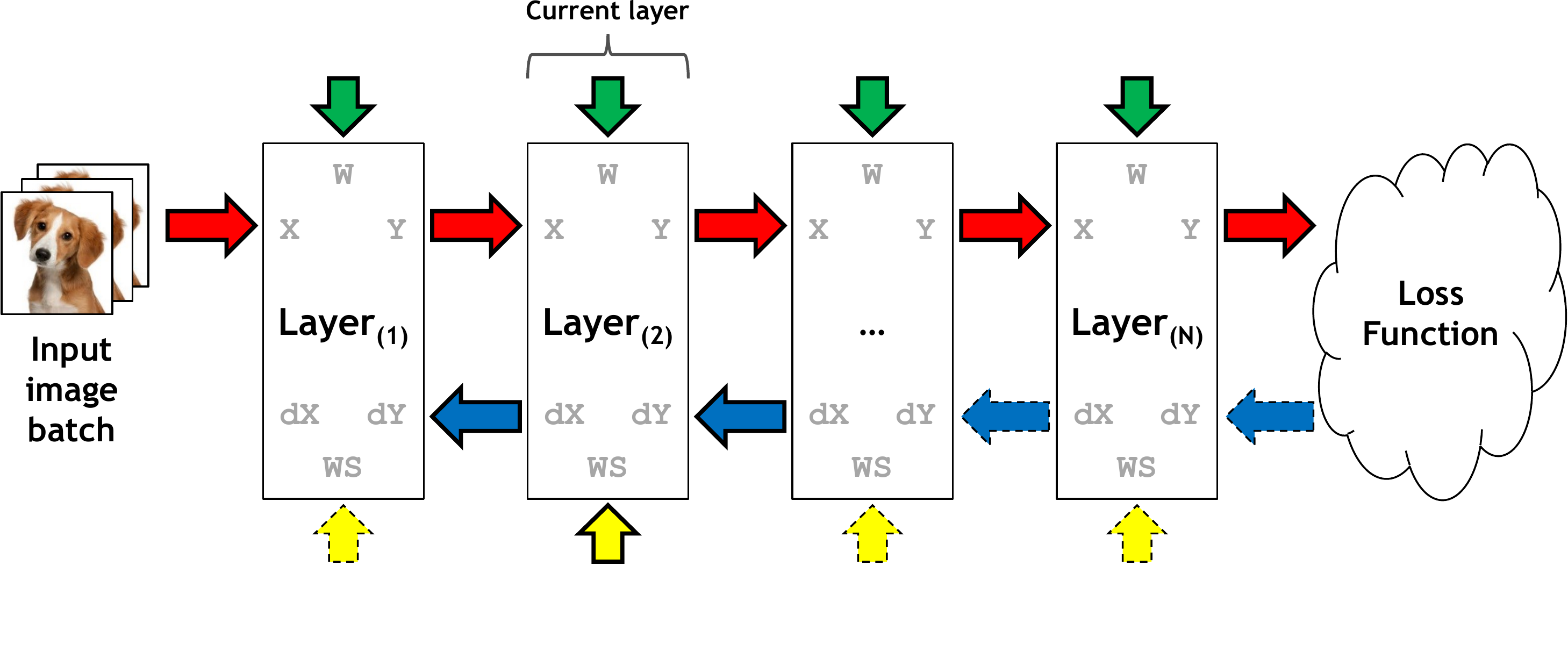}
\caption{Memory allocations required for linear networks using the
baseline memory manager (bold arrows). For inference, the sum of all green
(\weight) and red (\featureIn) arrows are allocated.  For training, two
additional data structures for \gradientOut and \gradientIn are required: both
are sized to the maximum of all blue (\gradientIn) arrows and are reused while
traversing back the layers during backward propagation. An optional temporary
buffer, called \emph{workspace} in cuDNN~\cite{cudnn_v4} (yellow arrow, \workspace), 
is needed in certain convolutional algorithms. The workspace buffer is 
sized with the maximum workspace requirement among all layers and is reused
during backward propagation.}
\vspace{-1em}
\label{fig:mem_allocation_policy_global} 
\end{figure}

\begin{figure}[t!] \centering
\includegraphics[width=0.48\textwidth]{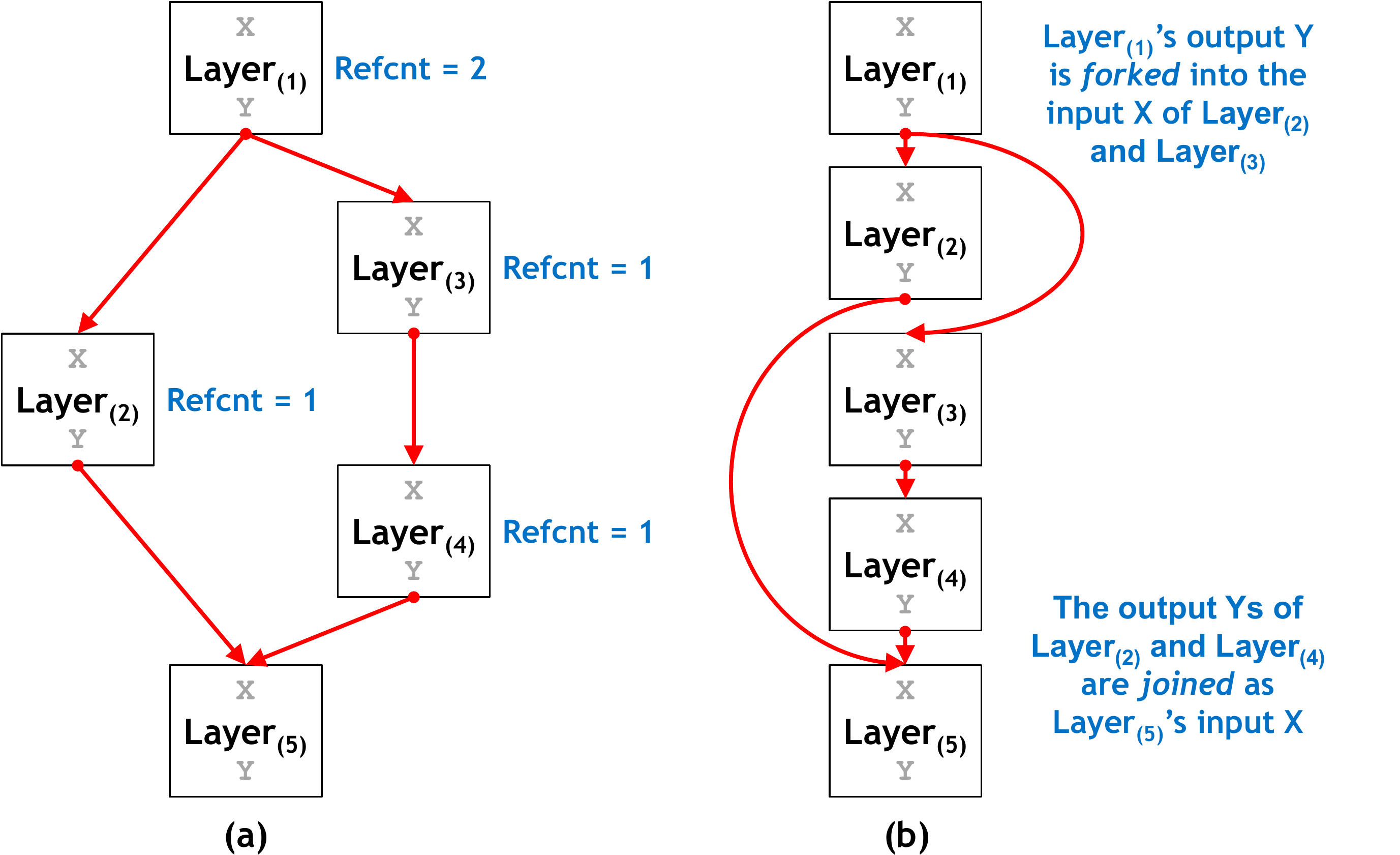}
\caption{
(a) The computation graph and its inter-layer dependencies of a GoogLeNet-style, non-linear feedforward network
during forward propagation. \texttt{Refcnt} refers to the number of consumer layers that 
depends on the current, producer layer's \featureOut. The order in which the GPU processes
each layer's forward computation is shown in (b), from \layer{1} to \layer{5}, highlighting the layer-wise computation
of DNN training. The producer-consumer relationship is reversed during backward propagation.
	}
\vspace{-1em}
\label{fig:nonlinear_dfg} 
\end{figure}

{\bf Forward Propagation.} Forward propagation is performed from the first
(input) layer to the last (output) layer, whereas backward propagation is
performed in the opposite direction (last to first layer), from right to left
in \fig{fig:mem_allocation_policy_global}.  Intuitively, forward propagation
traverses the network \emph{layer-wise} and performs the aforementioned feature
extraction and classification tasks on a given input, leading to an image
classification. During forward propagation, each layer applies a mathematical
operation to its input feature maps (\featureIn) and stores the results as
output feature maps (\featureOut). For linear feedforward DNNs, the
resulting \featureOut of \layer{n-1} is directly used as the input
\featureIn by \layer{n} (\fig{fig:mem_allocation_policy_global}). The
computation flow of forward propagation is therefore a \emph{serialized}
process, as \layer{n} can initiate its layer's operation only when the
preceding \layer{n-1} is finished with its computation and forwarded its output
\featureOut to \layer{n}'s input \featureIn.  Non-linear network
topologies can contain one-to-many (fork) and many-to-one (join) inter-layer
dependencies, but forward propagation still involves a series of layer-wise
computations as detailed in \fig{fig:nonlinear_dfg}. Note that the GPU can
only process a \emph{single} layer's computation at any given time due to such
inter-layer data dependencies. As a result, the minimum, per layer memory
allocations required are determined by the layer's input-output
relationships and its mathematical function\footnote{ Popular activation
functions (sigmoid/tanh/ReLU~\cite{alexnet}) can be refactored into an
\emph{in-place} algorithm using element-wise computation. Both Caffe and Torch
leverage this in-place memory optimization and only allocate memory space for
\featureOut and \gradientIn for forward (\featureOut) and backward (both
\featureOut and \gradientIn) propagation~\cite{torch_inplace}.  This paper
adopts this in-place optimization for both baseline and \vdnn for a
conservative evaluation.}. For instance, a CONV layer using the most
memory-efficient convolutional algorithm (e.g., implicit GEMM in
cuDNN~\cite{cudnn_v4}\footnote{ cuDNN (version 4.0) provides six different
convolutional algorithms. Implicit GEMM requires the least
memory allocation as no additional workspace is needed. FFT-based convolutional algorithms on
the other hand incur larger memory allocations because of the additional data
structures required to store the feature maps transformed into frequency
domain. More details are available in \cite{cudnn_v4,chetlur:2014:cudnn}.})
requires three data structures, the input/output feature maps (\featureIn and
\featureOut) and the weights of the layer (\weight) for forward propagation.
Employing a fast-fourier-transform (FFT) based convolution algorithm however
requires an additional, temporary workspace (\workspace) buffer to manage
transformed maps.

{\bf Backward Propagation.} For DNNs that are not fully trained, the inferred
image category might be incorrect.  As a result, a \emph{loss function} is used
to derive the magnitude of the inference error at the end of forward
propagation.  Specifically, the \emph{gradient} of the loss function is derived
with respect to the last \layer{N}'s output:

\begin{equation}
\frac{\partial Loss}{\partial Y_{(N)}}
\label{eqn:bwdprop}
\end{equation}

The value in \eqn{eqn:bwdprop} is forwarded to the last \layer{N} as its input
gradient maps (\gradientIn), and the output gradient maps (\gradientOut) are
derived based on the \emph{chain rule}~\cite{lecun_gd}:

\begin{equation}
\frac{\partial Loss}{\partial X_{(N)}} = \frac{\partial Loss}{\partial Y_{(N)}} \cdot \frac{\partial Y_{(N)}}{\partial X_{(N)}}
\label{eqn:chain_rule}
\end{equation}

Because the output \gradientOut ($\frac{\partial Loss}{\partial X_{(N)}}$) is
the product of the input \gradientIn ($\frac{\partial Loss}{\partial Y_{(N)}}$)
with $\frac{\partial Y_{(N)}}{\partial X_{(N)}}$, deriving the value of
\gradientOut for \layer{N} generally requires memory for both its input/output
gradient maps (\gradientIn and \gradientOut) and also the input/output feature
maps (\featureIn and \featureOut) for this layer. For linear 
networks, the calculated \gradientOut of \layer{N} is directly passed on to the
preceding \layer{N-1} to be used as \gradientIn for \layer{N-1}'s \gradientOut
derivation (\fig{fig:mem_allocation_policy_global}). This chain rule is similarly
used to derive the gradients of the weights to update the network model.

Similar to forward propagation, backward propagation is also performed
\emph{layer-wise} to the respective incoming gradient maps, \gradientIns.  Once
backward propagation reaches the first layer, the weights are adjusted using
the weight gradients so that the prediction error is reduced for the next
classification task. Hence, \emph{training} a network involves both forward and
backward propagation, which are repeated for millions to billions of
iterations. Because of the stochastic nature of SGD-based backward propagation,
the network input is generally batched with hundreds of images (e.g., $128$ and
$256$ images for best performing AlexNet and VGG-16), which increases 
memory allocation size but helps the network model better converge to an optimal
solution.

\subsection{Motivation: Scalable and Memory-Efficient DNN Design}
\label{sect:motivation}

To aid the design and deployment of neural networks, a variety of ML frameworks
have been developed in recent years, including Caffe, Torch, Neon, TensorFlow, and
Theano~\cite{bahrampour:2016:arxiv}.  The rich set of features offered by these
frameworks coupled with their ability to accelerate DNN training and inference using GPUs 
greatly simplifies the process of implementing neural
networks. Despite their flexibility, popular ML
frameworks suffer from severe limitations in the way they allocate and manage
memory. 

\begin{figure}[t!] 
\centering
\includegraphics[width=0.48\textwidth]{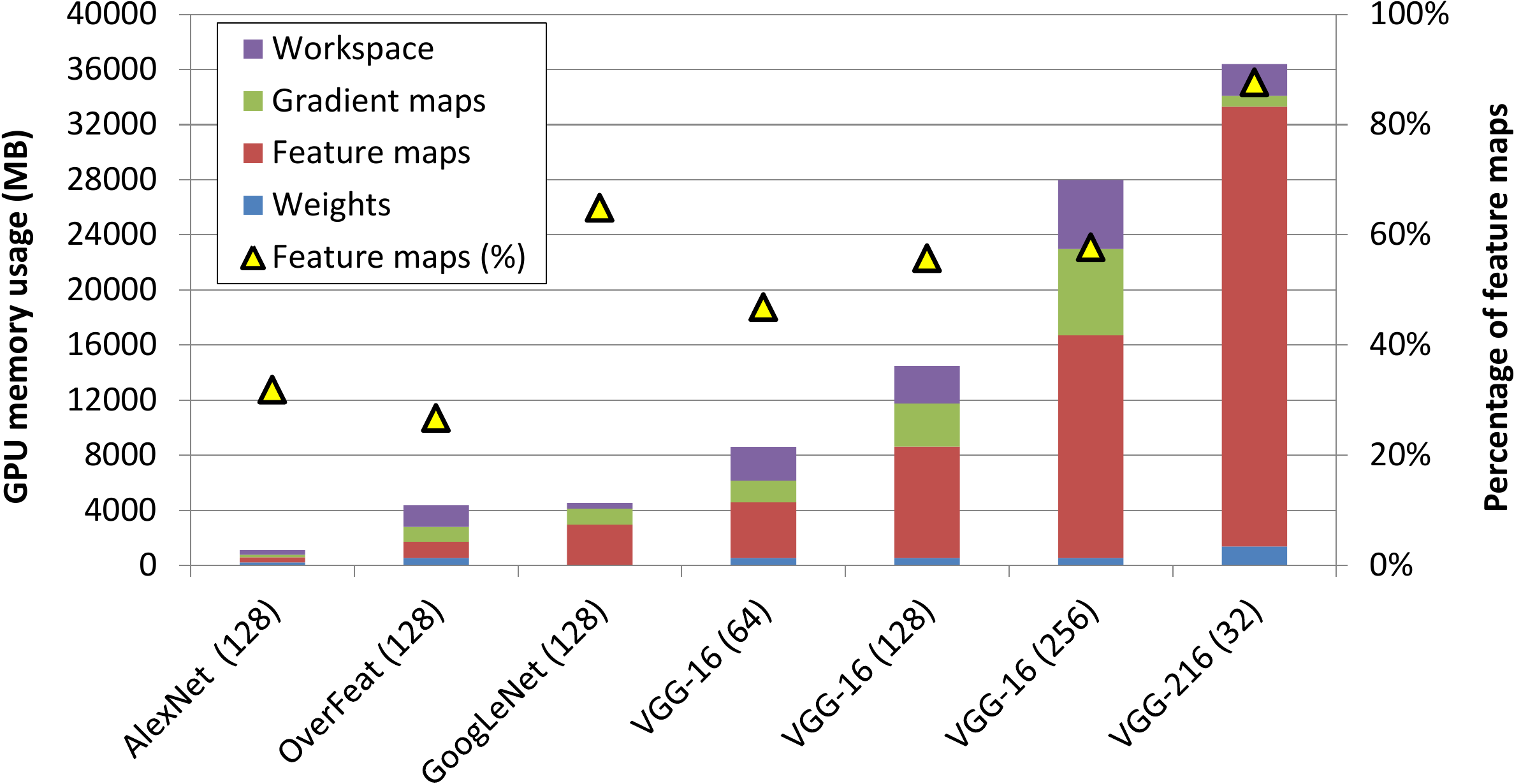}
\caption{Breakdown of GPU memory usage based on its functionality (left axis).
The right axis shows the fraction of allocated memory consumed by feature maps.
}
\vspace{-1em}
\label{fig:mem_usage_categorization} 
\end{figure}

\begin{figure}[t!]
\centering
\includegraphics[width=0.48\textwidth]{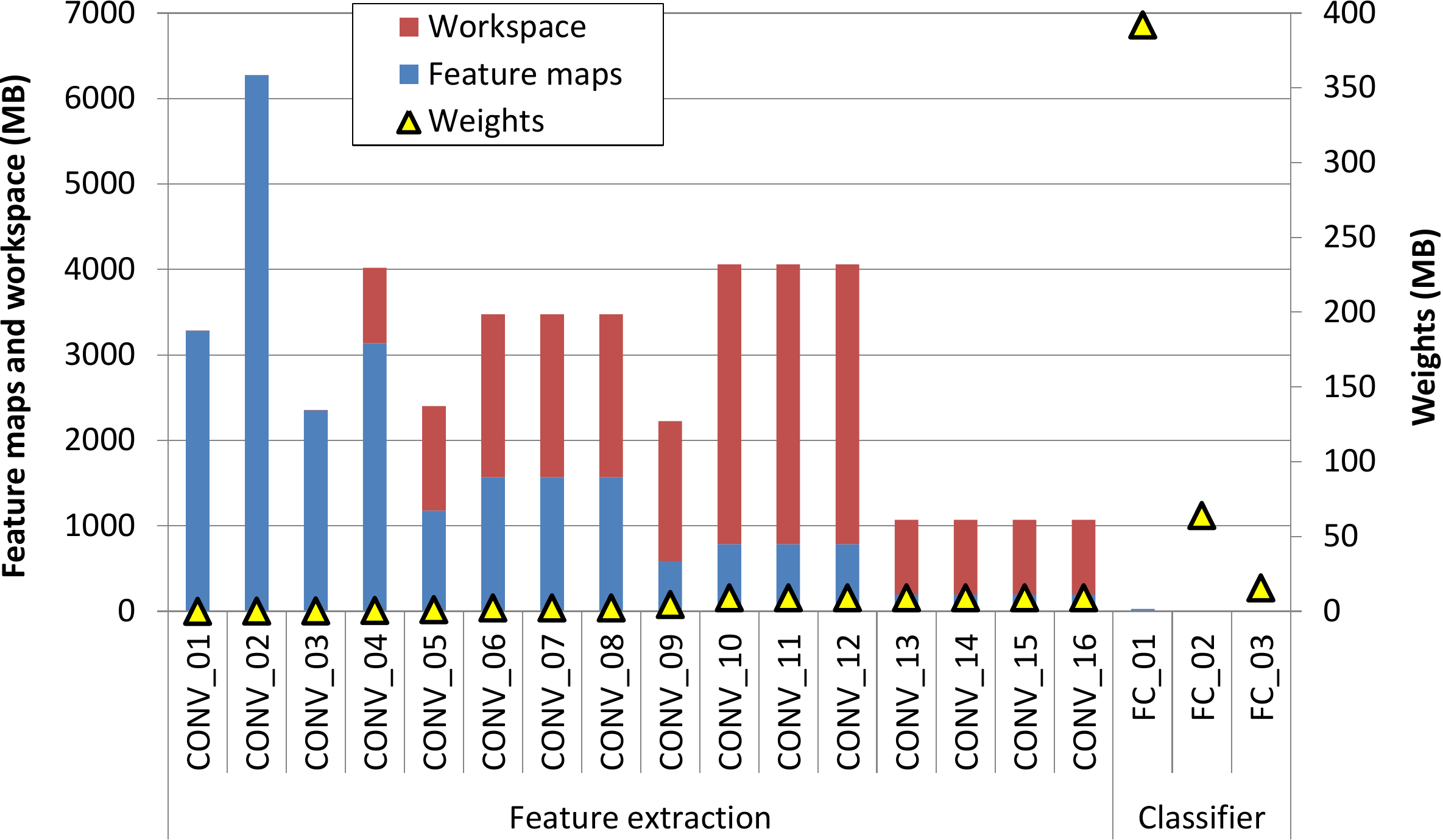}
\caption{Per layer memory usage of VGG-16~(256). For brevity, we only show the memory usage
during forward propagation and for layers that contain weights (CONV and FC). Left axis
corresponds to the sum of workspace and per layer input/output feature maps. The right axis corresponds to 
the memory consumption for storing weights. The memory usage during backward
propagation follows similar trends to this figure.}
\vspace{-1em}
\label{fig:memory_usage_per_layer}
\end{figure}

To illustrate the shortcomings of ML frameworks in managing memory, consider the example shown in
\fig{fig:mem_allocation_policy_global}. When training a DNN using existing ML
frameworks, the memory required across \emph{all} of the layers of the network
must fit within the physical GPU memory capacity.  The key reason for this
GPU-side, network-wide memory allocation strategy is to reap
performance benefits. More specifically, page-migration based virtualization
solutions that expose both CPU and GPU memory for page allocations (regardless
of whether the virtualization feature is provided by future CUDA runtime
extensions or programming models such as OpenMP (4.0)~\cite{openmp_4}) must
transfer pages via PCIe, which involves several latency-intensive processes
such as CPU interrupts for system calls, page-table updates, TLB
updates/shootdowns, and the actual page transfer.  Prior work~\cite{gpu_paging}
reported that the latency to page-in a single $4$ KB page to the GPU is
$20$ to $50$ $\mu$s, meaning the PCIe bandwidth utilization using
page-migration is $80$ to $200$ MB/sec, as opposed to the DMA
initiated \texttt{cudaMemcpy} that achieves an average $12.8$ GB/sec out of the
$16$ GB/sec maximum PCIe bandwidth. As the amount of data to be paged in/out 
via PCIe can be 10s of GBs for very deep networks (\fig{fig:vdnn_case_study_deeper_vgg}), 
ML frameworks will suffer from huge performance penalties when relying on page-migration
for training DNNs.

Note that because of the layer-wise gradient update rule of the backward
propagation algorithm (property of the chain rule, \sect{sect:dnn_training}),
each layer's feature maps (\featureIn) are later \emph{reused} during its own
backward propagation pass. This means that \emph{all} \featureIns must
still be available in GPU memory until backward computation is completed.
\fig{fig:mem_usage_categorization} shows the amount of memory usage based on
its functionality and the growing significance of feature maps as networks
become deeper. Because deeper networks need to keep track of a larger number of
\featureIns, the fraction of memory allocated for feature maps grows
monotonically as the number of layers increases. Training the network
itself is still done layer-wise, however, regardless of the depth of the neural network.
The baseline network-wide memory allocation policy is therefore both extremely
wasteful and not scalable because it does not take into account the layer-wise
DNN training.  \fig{fig:memory_usage_per_layer} shows the per layer memory
usage of VGG-16 during forward propagation, which provides the following key
observations.  First, the intermediate feature maps and workspace (left axis)
incur an order of magnitude higher memory usage compared to the weights (right
axis) of each layer. Second, most of these intermediate data structures are
concentrated on the feature extraction layers and are less significant in the
later classifier layers.  Third, the weights, while smaller in size compared to
these intermediate data, are mostly concentrated on the classifier layers due
to their full connectivity.  Lastly, the per layer memory usage is much smaller
than the $28$ GB of memory required by the baseline policy
(\fig{fig:mem_usage_global_vs_local}), showing significant opportunities for
memory savings with a fine-grained, layer-wise memory management policy.

%% file: tex/vdnn.tex
\section{Virtualized DNN}
\label{sect:vdnn}

The design objective of our virtualized DNN (\vdnn) memory manager is to
\emph{virtualize} the memory usage of DNNs, using both GPU and CPU memory,
while minimizing its impact on performance.  \vdnn is completely transparent to
	the programmer as the allocation, placement, movement, and release of data is
	seamlessly orchestrated by the system architecture and the runtime system.
	Such abstraction enables ML practitioners to focus more on their ML algorithm
	and not have to worry about the low level details of GPU memory management.
	\vdnn primarily optimizes the memory usage of the feature extraction layers
	as the majority of memory usage is concentrated on these
	layers, accounting for 81\% of memory usage on AlexNet and 96\% on
	VGG-16~(256). More specifically, we target the feature maps of these feature
	extraction layers as these intermediate data structures account for the
	majority of GPU memory usage (\fig{fig:mem_usage_categorization} and
	\fig{fig:memory_usage_per_layer}).  The intuitions of \vdnn can also be
	applied to weights and to the classification layers, but with less of a
	memory saving benefit.

\begin{figure}[t!] \centering
\includegraphics[width=0.48\textwidth]{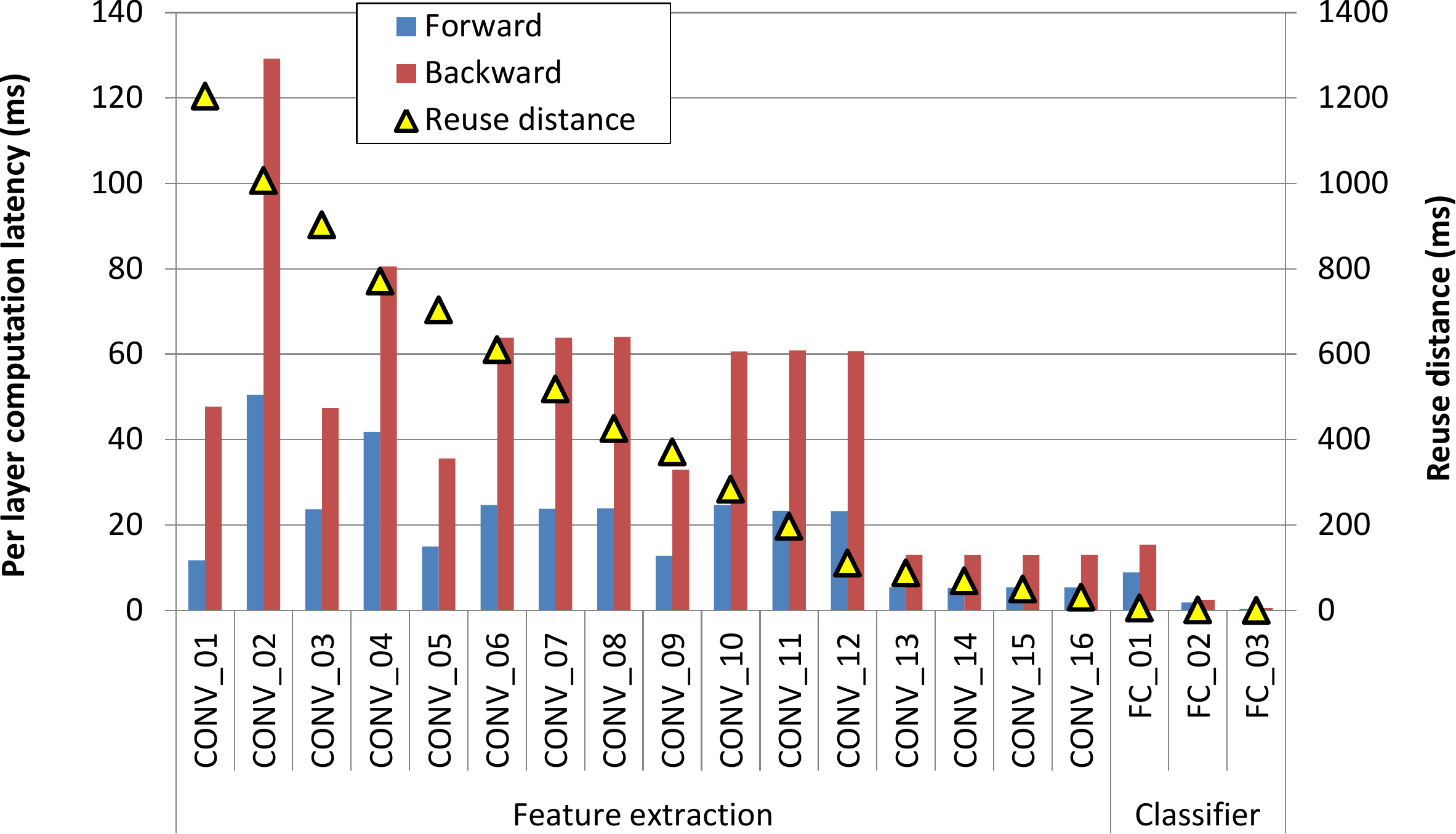}
\caption{
VGG-16's per layer computation latency for forward and backward propagation (left axis). Right axis
shows the reuse distance of each layer's input feature maps, \featureIn. We define the reuse distance of a
\layer{n}'s \featureIn as the latency between the completion of \layer{n}'s forward propagation and the start of the same \layer{n}'s backward
propagation.}
\vspace{-1em}
\label{fig:reuse_distance}
\end{figure}

\subsection{Design Principle}
\label{sect:vdnn_principle}

Previous sections highlighted the fact that the memory requirement per individual layer is
substantially smaller than what is actually provisioned with the baseline,
\emph{network-wide} memory allocation policy.  \vdnn adopts a sliding-window
based, \emph{layer-wise} memory management strategy in which the runtime memory
manager conservatively allocates memory from its memory pool for the immediate
usage of the layer that is currently being processed by the GPU.  Intermediate
data structures that are not needed by the current layer are targeted for
memory release to reduce memory usage.

{\bf Forward Propagation.}  As discussed in \sect{sect:motivation}, deep networks have to keep
track of a large number of the intermediate feature maps (\featureIns) that are
extracted during forward propagation. Once a given \layer{n}'s forward
computation is complete, however, \layer{n}'s \featureIn is not reused until
the GPU comes back to the same \layer{n}'s  corresponding backward computation.
Because the reuse distance of \layer{n}'s \featureIn is on the order of milliseconds to
seconds (e.g., more than $60$ ms and $1200$ ms for the first layer of AlexNet
and VGG-16~(64), respectively), deep networks end up allocating a significant
number of \featureIns that effectively camp inside the GPU memory without
immediate usage (\fig{fig:reuse_distance}).  As a result, tackling these \featureIns for memory
optimization is crucial for efficient utilization of GPU memory as these
intermediate data account for a significant fraction of memory
allocations (\fig{fig:mem_usage_categorization}).  \vdnn therefore
conditionally \emph{offloads} these intermediate \featureIns to CPU
memory via the system interconnect (e.g., PCIe, NVLINK~\cite{nvlink}) if they are targeted for memory release.
\sect{sect:vdnn_transfer_policy} details the \vdnn memory transfer policy that
decides which layers are chosen for offloading its \featureIn. Once the offload
operation is complete, \vdnn \emph{releases} the offloaded \featureIn from the
memory pool to reduce GPU memory usage.

\begin{figure}[t!] \centering
\includegraphics[width=0.48\textwidth]{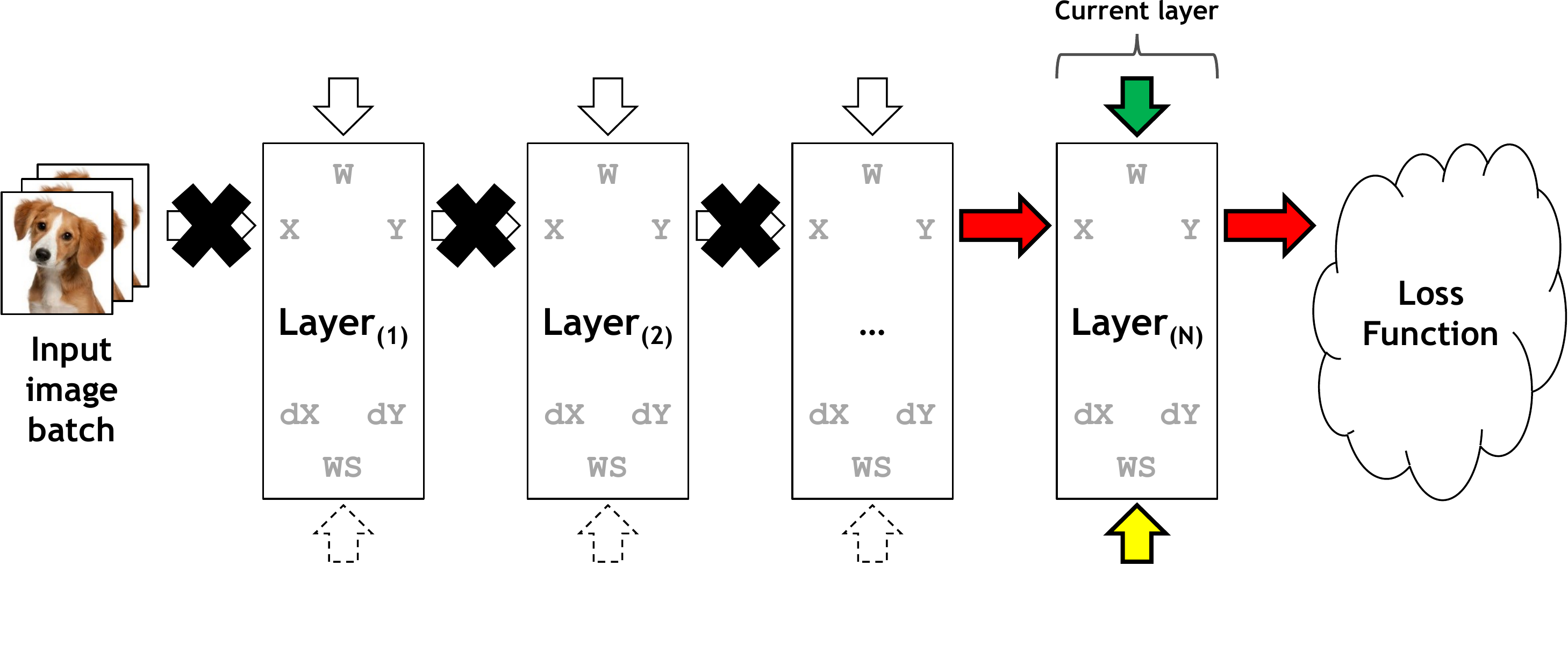}
\caption{
Execution flow of a linear network during forward propagation. The figure assumes that \layer{N}
is currently being processed by the GPU. During this layer's forward computation, 
the data associated with the arrows marked with
black $X$s (all preceding layer's input feature maps) are not used and can safely be released from
the memory pool.
}
\vspace{-1em}
\label{fig:mem_allocation_policy_local_fwdprop} 
\end{figure}

\begin{figure}[t] \centering
\includegraphics[width=0.48\textwidth]{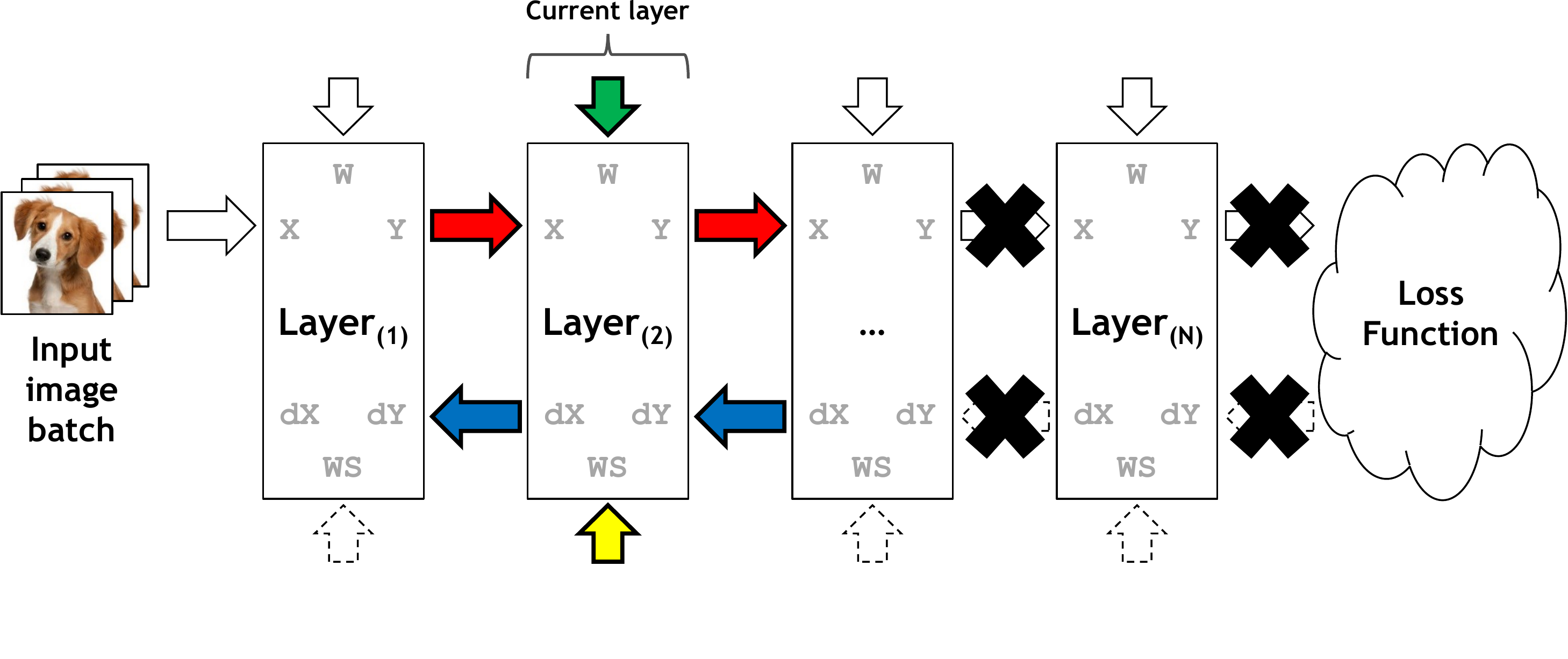}
\caption{Execution flow of a linear network during backward propagation. The figure assumes that
\layer{2} is currently being processed by the GPU. Data associated with the 
arrows marked with black $X$s can safely be released because they will not be reused
during the training of this input image batch.}
\vspace{-1em}
\label{fig:mem_allocation_policy_local_bwdprop} \end{figure}

Care must be taken however when evaluating the feasibility of offloading
a layer's input \featureIn. This is because, for non-linear network topologies,
multiple layers can be the consumers of a previously computed layer's output
feature maps (\featureOut). For instance, \layer{2} and \layer{3} in
\fig{fig:nonlinear_dfg} are both using the output \featureOut of \layer{1} as
its input \featureIn.  Offloading and consequently releasing the input
\featureIn of \layer{2}, before reaching \layer{3}'s forward computation, is
problematic as these two layers share the same data structure for the input
\featureIn.  \vdnn therefore keeps track of the inter-layer 
dependencies in the form of a dataflow graph (e.g., \texttt{Refcnt}
in \fig{fig:nonlinear_dfg}) and allows the offload/release operation to be initiated 
only when the currently processing layer is the last consumer of its input feature maps.
\fig{fig:mem_allocation_policy_local_fwdprop} is an example execution flow of a linear DNN
during forward propagation, highlighting when it becomes safe to release a layer's \featureIn.

{\bf Backward Propagation.} Similar to forward propagation, \vdnn
aggressively releases data structures that are not needed for training the
remaining layers' backward computation. During \layer{n}'s backward
propagation, \layer{n+1}'s \featureOut and \gradientIn are no longer required
because the GPU has already completed the gradient updates for this layer
(\fig{fig:mem_allocation_policy_local_bwdprop}).  Again, by leveraging the
layer-wise DNN backward propagation, \vdnn immediately frees up a layer's
\featureOut and \gradientIn once this layer's backward computation is complete.
\featureIn and \gradientOut are not released as the preceding layer's
backward propagation will be needing these values for gradient derivation.  Note
that if a layer has offloaded its \featureIn to host memory, \vdnn should guarantee
that the offloaded data is copied back to GPU memory before the gradient update
is initiated.  Naively copying back the data on-demand will serialize the
backward computation behind the memory copying operation of \featureIn. 
\vdnn therefore launches a \emph{prefetch} operation for
\layer{n}'s offloaded feature maps, which is overlapped with
\layer{m}'s backward computation, with $n<m$, so that prefetching is launched
before its actual usage, hiding prefetching latency.

\begin{figure}[t] \centering
\includegraphics[width=0.48\textwidth]{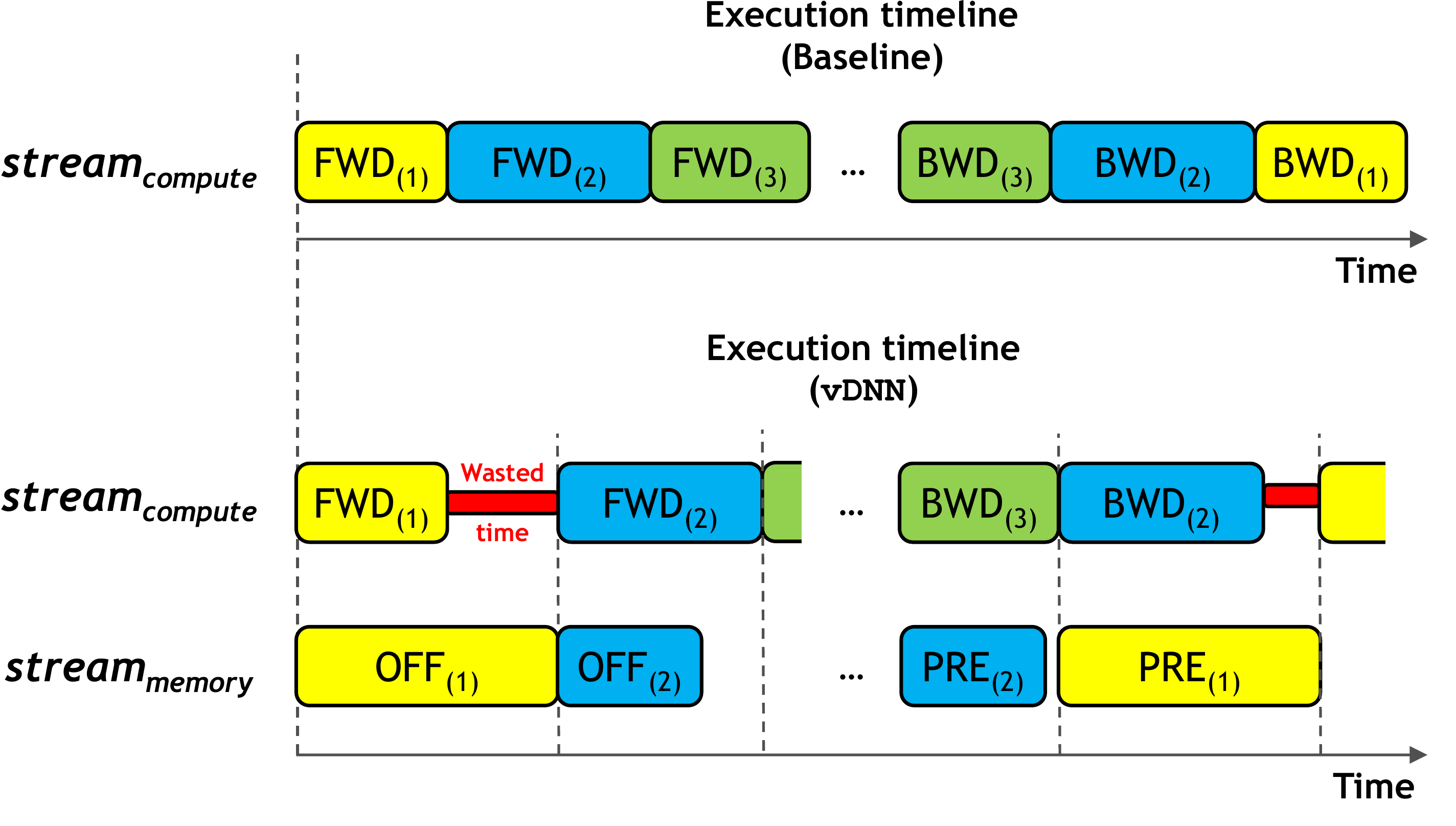}
\caption{Performance effect of offload and prefetch.  FWD$_{(n)}$ and
BWD$_{(n)}$ are the forward and backward computations for
\layer{n}, respectively. OFF$_{(n)}$ is the offloading of \layer{n}'s
\featureIn and PRE$_{(n)}$ is the corresponding prefetch operation for
\layer{n}.} 
\vspace{-1em}
\label{fig:vdnn_offload_prefetch_mechanism} \end{figure}

\subsection{Core Operations And Its Design}
\label{sect:vdnn_core_op}

\vdnn is prototyped as a layer on top of cuDNN~\cite{cudnn_v4}.  
Each layer keeps track of the cross-layer data dependencies of input/output feature maps so
that the \vdnn offload and release operations are properly scheduled.
\vdnn employs two separate CUDA streams~\cite{cuda} to
overlap normal DNN computations with the memory allocation, movement, and
release operations of \vdnn.  \streamCompute is the CUDA stream that interfaces
to the cuDNN handle and sequences all the layer's forward and backward
computations. \streamMemory manages the three key components of \vdnn; the
memory allocation/release, offload, and prefetch.

{\bf Memory Allocation/Release.} The CUDA library only supports
\emph{synchronous} memory (de)allocations, meaning that any calls to
\texttt{cudaMalloc()} or \texttt{cudaFree()} will enforce an additional
synchronization across all the GPUs within a node. 
To safely enable \vdnn memory operations while not fall into the pitfalls
of synchronous CUDA APIs, we employ the open-source \emph{asynchronous} memory
allocation/release API library distributed by NVIDIA~\cite{cnmem}. When the program
launches, the \vdnn memory manager is allocated with a memory pool that is
sized to the physical GPU memory capacity.  Whenever \vdnn allocates (and
releases) data structures, the underlying memory manager will reserve (and
free) memory regions from this memory pool without having to call
\texttt{cudaMalloc()} or \texttt{cudaFree()}.

\begin{figure}[t!] \centering
\includegraphics[width=0.48\textwidth]{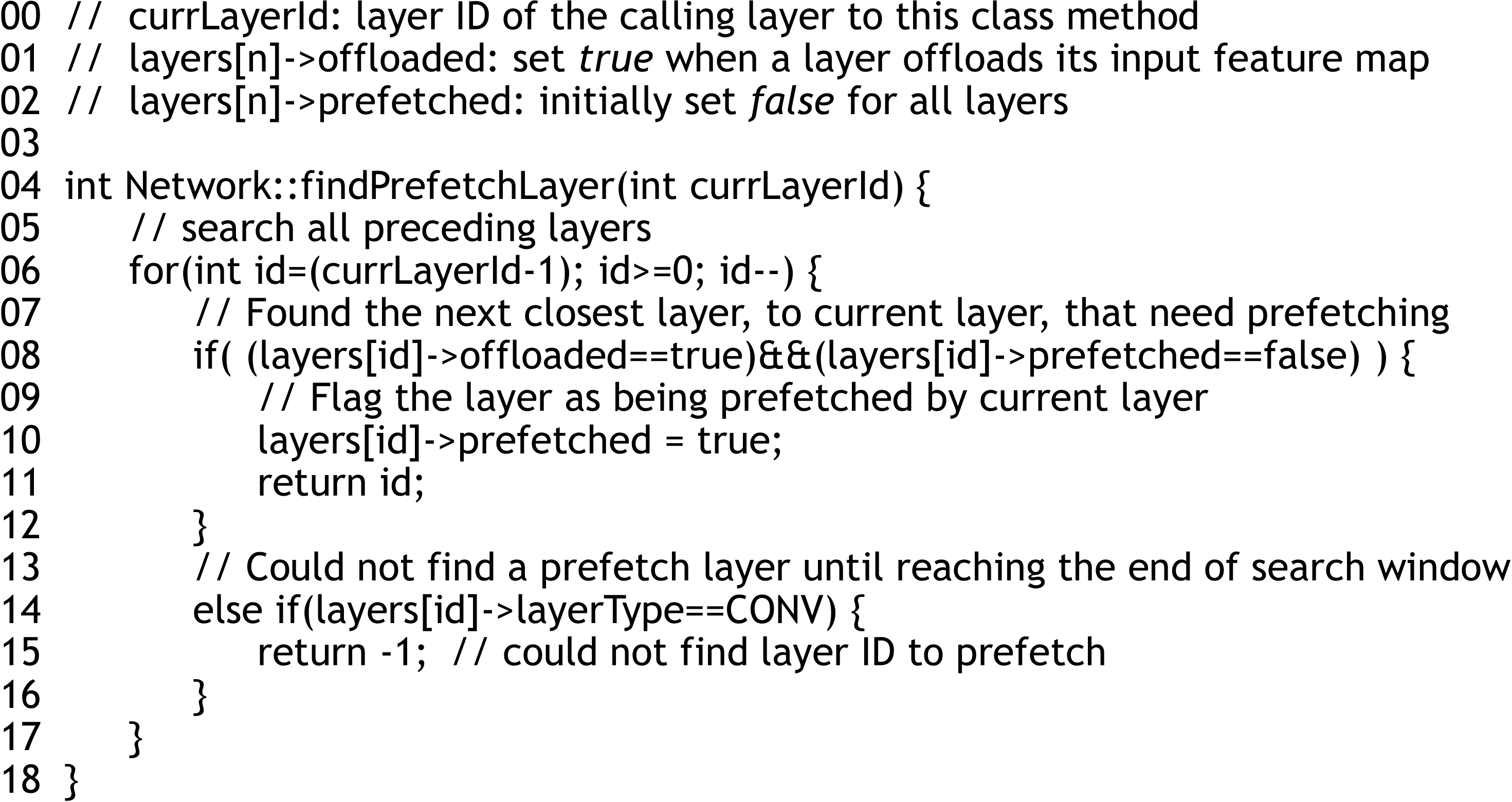} \caption{Pseudo
code explaining how \vdnn finds layers to prefetch. Notice how the search
operation is only up to the next closest CONV layer (line 14), guaranteeing
that the prefetched \featureIn will not end up being used too far away in the
future as it restricts the prefetch layer to be within the search window of
layers.} 
\vspace{-1em}
\label{fig:code_prefetch} \end{figure}

{\bf Memory Offload.} Offloading input feature maps 
is one of the key enablers of \vdnn's memory savings. When a layer
is chosen for offloading, \vdnn first allocates a pinned host-side
memory region using \texttt{cudaMallocHost()}. \streamMemory then launches a
non-blocking memory transfer of this layer's \featureIn to the pinned memory 
via PCIe using \texttt{cudaMemcpyAsync()}, overlapping it with the same layer's
forward computation of cuDNN. The current implementation of \vdnn synchronizes
\streamCompute and \streamMemory at the end of each layer's forward computation
if \streamMemory has offloaded its feature maps. This approach guarantees that the
offloaded data is safely released from the memory pool before the next layer
begins forward computation, maximizing the memory saving benefits of 
offloading. Because the \featureIns of CONV and POOL layers are read-only data
structures, overlapping \layer{n}'s offload operation with the same layer's
forward propagation does not create any correctness issues.  ACTV
layers are already refactored into an in-place algorithm and only use 
\featureOut and \gradientIn for gradient updates, obviating the need for memory
offloading (\sect{sect:dnn_training}).
\fig{fig:vdnn_offload_prefetch_mechanism} provides an overview of
\vdnn's offload operation.  Here, the baseline system is able to immediately
launch \layer{2}'s forward computation once \layer{1} is complete. The
execution of \layer{2} is stalled for \vdnn, because \streamCompute
must wait until the offloading operation of \streamMemory is complete, blocking
\layer{2}'s computation. The computation of \layer{3} is not delayed however
because the offload latency for \layer{2} is completely hidden inside the latency
to compute the same layer's forward propagation.

{\bf Memory Prefetch.} Similar to offloading, prefetching the
offloaded \featureIns back to GPU memory is implemented using
\texttt{cudaMemcpyAsync()} to overlap data transfers with the computations of
backward propagation.  However \streamMemory launches prefetch operations
in the reverse order relative to the offload operations from forward propagation (\fig{fig:vdnn_offload_prefetch_mechanism}). As mentioned
in \sect{sect:vdnn_principle}, the general rule of prefetching is to overlap
the memory copy operation of \layer{n}'s offloaded data with \layer{m}'s
backward computation, with layer ID $m$ always being higher than $n$ to
maximize the benefit of both prefetching and latency hiding.  In other words,
when the GPU starts the backward propagation of \layer{m}, \vdnn 
determines the best layer to prefetch among the preceding layers (as $n<m$).

If the distance between the prefetched \layer{n} and overlapping
\layer{m} is too far away, the memory saving benefit of \vdnn offloading will
be reduced because the reuse time of this prefetched data will be distant in
the future.  In other words, prefetching data too early in time will again
suboptimally utilize GPU memory as the prefetched data will once again
camp inside the GPU memory without immediate usage. We carefully
designed the \vdnn prefetch algorithm to avoid this pitfall and balance the
memory saving benefits of offloading with the timeliness of prefetching.
\fig{fig:code_prefetch} is a pseudo-code of the \vdnn prefetch algorithm that
determines the best candidate layer for prefetching. Before \streamCompute
starts a layer's backward computation, \vdnn first searches for a potential
layer that requires prefetching of its \featureIn.  If the search operation is
successful (line 11), the layer ID to be prefetched is returned by
\texttt{findPrefetchLayer} routine and is used to launch its prefetch
operation via \streamMemory.  Similar to offloading, \vdnn
synchronizes \streamCompute and \streamMemory so that
the next layer's backward computation is stalled until the prefetch operation
is finalized. Consequently, any prefetch operation launched during
\layer{n}'s backward computation is guaranteed to be ready before \layer{n-1}'s
computation. This benefit of course comes at the cost of a potential performance loss
when the prefetch latency is longer than the overlapped computation,
which we detail in \sect{sect:perf}.

\subsection{vDNN Memory Transfer Policy}
\label{sect:vdnn_transfer_policy}

Determining the best layers to offload their feature maps is a
multi-dimensional optimization problem that must consider: 1) GPU
memory capacity, 2) the convolutional algorithms used and the overall layer-wise
memory usage, and 3) the network-wide performance. The first two factors
determine whether we are able to train the network at all (which we
refer to as \emph{trainability} of a network), while the last factor
decides overall training \emph{productivity}. If \vdnn were to use the
most memory-efficient algorithm for all layers (e.g., implicit
GEMM in cuDNN~\cite{cudnn_v4} which does not require any \workspace allocations) while also
having all layers offload/prefetch, the GPU memory usage will be the lowest.
Performance will likely suffer, however, compared to a baseline with the
fastest convolutional algorithms adopted for each layers; the performance loss
primarily comes from 1) the additional latency possibly incurred due to
offload/prefetch, and 2) the performance difference between memory-optimal
implicit GEMM and the performance-optimal convolutional algorithm.  Going with
the fastest algorithm, without any offload/prefetch, will result
in the highest possible performance, but the potential memory overheads for the
faster algorithm's workspace and the cumulative \featureIns that camp inside
the GPU memory will likely overflow GPU memory.  Given that
optimizing the layer-wise memory usage and its performance is in itself
a multi-dimensional optimization problem, selecting the most optimal
hyperparameters across the entire network is non-trivial. We therefore adopt
the following heuristic-based memory transfer policies that narrow 
the parameter choices and simplify the optimization problem, while still performing
robustly in practice.

{\bf Static vDNN.} Feature extraction layers are mostly composed of
CONV and ACTV layers with intermittent POOL layers that downsize the
dimensionality of the feature maps. More than 70\% to 80\% of the
(forward/backward) computation time however is spent on the CONV layers for
deep neural networks. We therefore evaluate two \emph{static} \vdnn memory
transfer options that exploit this computational characteristic. The first
option we explore is to have the \vdnn memory manager offload all of the \featureIns
of all of the layers. This policy, \vdnnAll, is our most
memory-efficient solution as all \featureIns are offloaded and released from
the GPU, drastically reducing device memory usage.  The second \vdnn policy is
to only offload \featureIns for the CONV layers and leave the remaining
layers' \featureIns resident inside GPU memory (\vdnnConv). The \vdnnConv
policy is based on the observation that CONV layers have a much longer
computation latency than ACTV/POOL layers, being more likely to effectively hide the
latency of offload/prefetch. Not surprisingly the performance of \vdnnConv is
generally higher than \vdnnAll. But \vdnnAll has the advantage of consuming
the least GPU memory, significantly enhancing the trainability of a
DNN.  We later evaluate the memory usage and performance of
these two static policies with both the memory-optimal and performance-optimal
convolutional algorithms employed.

{\bf Dynamic vDNN.} While static \vdnn is simple and easy to implement,
it does not account for the system architectural components that determine the
trainability and performance of a DNN (e.g., maximum compute FLOPs and
memory bandwidth, memory size, effective PCIe bandwidth, etc). For DNNs that
comfortably fit within GPU memory, neither \vdnnAll nor
\vdnnConv is optimal as the best approach is to have all the memory allocations
resident in GPU without any offloading and employ the fastest possible
convolutional algorithm.  Large and deep networks, on the other hand, might not
have the luxury of using faster convolutional algorithm.  So being able to fit
such network on the GPU is the best optimization \vdnn could make.  We
therefore develop a \emph{dynamic} \vdnn policy that automatically determines
the offloading layers and the convolutional algorithms employed, at runtime, to
balance the trainability and performance of a DNN. Dynamic \vdnn leverages
several properties of DNN training. First, we exploit the 
millions to billions of iterations of the same forward/backward propagation pass
that are required for training.  NVIDIA's cuDNN provides a runtime API that
experiments with all available convolution algorithms for a given layer,
evaluating each algorithm's performance and its memory usage.  Current ML
frameworks leverage this API to undergo an initial profiling stage to
determine the best algorithms to deploy for each CONV layer for best
performance.  The overhead of such profiling is on the order of a few tens of
seconds, which is negligible relative to the days to weeks required for DNN training.
Our dynamic \vdnn augments this profiling
stage with a number of additional profiling passes 
to select the best layers to offload and the best per layer algorithm.
Once the baseline profile stage is completed
and the fastest possible convolutional algorithms are derived for all CONV
layers, dynamic \vdnn employs the following additional profiling passes:

\begin{enumerate}

\item First, the static \vdnnAll is tested for a single training pass with all
CONV layers using the memory-optimal, no-\workspace incurred algorithm. 
This initial pass determines if the target DNN can be trained at all as
it requires the least GPU memory.

\item If \vdnnAll passed, another training phase is launched with all CONV
layers employing the fastest algorithms but without any offloading. Such
a configuration, if it passes successfully, will be adopted for the rest of the full training procedure
as it provides the highest performance while guaranteeing trainability.
If this profiling phase fails due to memory oversubscription, two additional training
passes are tested with the same fastest algorithms, but with \vdnn
offloading enabled for both \vdnnConv and \vdnnAll respectively. If successful, \vdnn employs the
succeeded configuration for the rest of training. If both \vdnnConv and \vdnnAll fails, we
move on to the next profiling pass below to further reduce memory usage.

\item The last phase is based on a \emph{greedy} algorithm that tries to
locally reduce a layer's memory usage, seeking a global optimum state in terms
of trainability and performance.  When traversing through each layer, \vdnn
first calculates whether using the fastest algorithm will overflow the GPU
memory budget. If so, then the given layer's convolutional algorithm will be
\emph{locally} downgraded into a less performant but more memory-efficient one,
until it reaches the memory-optimal implicit GEMM.  This greedy-based approach
first tries \vdnnConv with each CONV layer initially using its own
performance-optimal algorithm. If \vdnnConv fails, then another training pass
is launched with the more memory-efficient \vdnnAll. If \vdnnAll also fails
with this greedy algorithm, then \vdnn resorts back to the very first \vdnnAll
solution, with the memory-optimal, no-\workspace algorithms applied across the
entire network. 

\end{enumerate}

While other possible settings might better balance performance
and trainability, we find that our dynamic \vdnn performs competitively
without having to exhaustively search for globally optimal parameter selections.

%% file: tex/eval.tex
\section{Methodology}

\subsection{vDNN Memory Manager} 
\label{sect:method_vdnn}

We implemented a host-side memory manager that interacts with the latest and
fastest version of cuDNN 4.0~\cite{cudnn_v4}, serving as the GPU back-end. All
the layers that constitute a DNN's feature extraction layer 
have been implemented using cuDNN, and the execution of each layer is
orchestrated using two CUDA streams, \streamCompute and \streamMemory as
discussed in \sect{sect:vdnn_core_op}. The classification layers remain
unchanged and use the same cuBLAS routines used in Torch. The \vdnn API closely
resembles that of Torch and Caffe, providing the high level abstractions of the
target DNN and each of its layer compositions.

While there are subtle differences between Torch, Caffe, and Theano's memory
allocation scheme, prior work~\cite{bahrampour:2016:arxiv} quantitatively
demonstrated that all three frameworks exhibit comparable performance and
memory requirements\footnote{ Because TensorFlow is the least
performant in terms of GPU memory usage and training speed~\cite{bahrampour:2016:arxiv}, we do not discuss its
memory management policy further in this paper.}.  We therefore choose Torch's
memory management policy as baseline to compare against \vdnn given
its widespread deployment across both academia and industry (e.g., Facebook and
Google DeepMind). This baseline policy adopts the
network-wide allocation policy discussed in \sect{sect:motivation}.  However,
we further improve this baseline policy using the following strategy to reduce memory
consumption during the backward propagation
phase~\cite{torch_blog_resnet,mxnet}: rather than allocating separate
\gradientIn and \gradientOut for all individual layers, we only allocate the minimally required number of 
each of these data structures and reuse them after each layer's backward
computation is complete (\fig{fig:mem_allocation_policy_global}).  

\begin{figure*}[t!]
\centering
\includegraphics[width=1.0\textwidth]{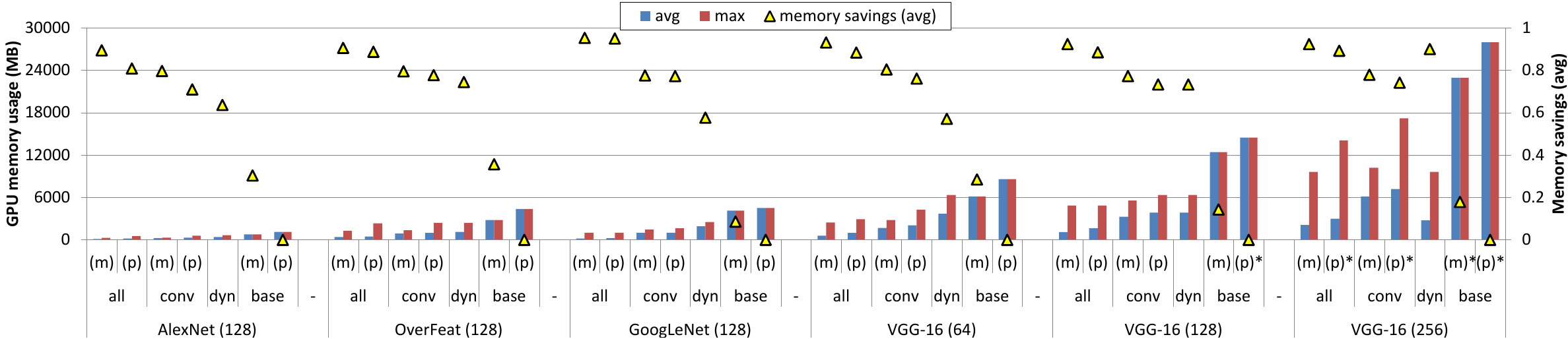}
\caption{Average and maximum memory usage (left axis). Right axis corresponds to the savings in average memory usage.}
\vspace{-1em}
\label{fig:vdnn_mem_usage_max_avg}
\end{figure*}

\subsection{GPU Node Topology}

We conducted experiments on NVIDIA's Titan X~\cite{titan_x}, which
provides the highest math throughput (single precision throughput of
$7$ TFLOPS), memory bandwidth (max $336$ GB/sec), and memory capacity ($12$ GB)
in the family of Maxwell GPUs. The GPU communicates with an Intel
i7-5930K (containing $64$ GB of DDR4 memory) via a PCIe switch (gen3),
which provides a maximum $16$ GB/sec data transfer bandwidth.

\subsection{DNN Benchmarks} 
\label{sect:benchmark}

{\bf Conventional DNNs.} First, we evaluate existing, state-of-the-art
ImageNet winning DNNs: AlexNet~\cite{alexnet}, OverFeat~\cite{overfeat}, GoogLeNet~\cite{googlenet},
and three different batch sizes for VGG-16 (the deepest network with 16 CONV and 3 FC
layers)~\cite{vggnet}. The network configurations of these DNNs (e.g., layer 
type, batch size, etc.) are
identical to the reference models maintained by the researchers at
Facebook~\cite{soumith:convnet_bench}. While the memory usage of AlexNet, 
OverFeat, and GoogLeNet is already below the $12$ GB memory capacity of Titan X
(\fig{fig:mem_usage_categorization}), we use it to evaluate the performance
regression on these networks with \vdnn. VGG-16 is one of the largest and
deepest DNN architecture to date, requiring substantial memory capacity for
trainability (using up to $28$ GB of memory when trained with the best
performing batch size of 256). Accordingly, Simonyan and Zisserman~\cite{vggnet} parallelized
VGG-16 (256) across four GPUs, with each GPU training VGG-16 (64) that fits within a single GPU memory
budget. We therefore study VGG-16 with three batch sizes ($64$/$128$/$256$) and use it as
a representative, future-looking DNN architecture that stresses the memory capacity limits of
today's GPUs.

{\bf Very Deep Networks.} To highlight \vdnn's scalability in
training very deep networks, we collected a second set of benchmarks by
extending the number of CONV layers of VGG, from 16 CONV layers to 416 CONV
layers.  The original VGG network features a homogeneous architecture that only
uses $3\times3$ convolutions (with stride 1 and pad 1) and $2\times2$ pooling
operations (with stride 2), from the first to the last feature extraction
layer. The feature extraction layers are conceptually divided into five groups
of CONV layers, separated by the intermediate POOL layers. The only difference
among these CONV layer groups is that the number of output feature maps grows
from $64$ to $512$, from the first to the last layer group. Simonyan and
Zisserman~\cite{vggnet} studied the effect of layer depth on classification
accuracy by incrementally adding more CONV layers to each of these layer
groups, going from 8 CONV layers to 16 CONV layers. We follow similar measures
to deepen the layer depth of VGG by gradually adding $100$ more CONV layers to
VGG-16, resulting in VGG-$116$/$216$/$316$/$416$ configurations.  
Each addition of $100$ CONV layers is done by adding $20$ more CONV layers to each of the
five CONV layer groups. The added CONV layers have the same number of output
feature maps that are employed for that layer group.  We use these four
VGG-style networks to perform a case study on \vdnn's scalability on training
very deep networks that require much more memory. Compared to conventional DNNs
whose input batch size is in the order of hundreds of images, we study these
very deep networks with a relatively small batch size of $32$ in order to
highlight the memory scaling effect of layer \emph{depth} on DNNs.

%% file: tex/results.tex
\section{Results} 
\label{sect:results}

This section evaluates the effect of \vdnn on GPU memory usage, off-chip memory
bandwidth utilization, GPU power consumption, and overall performance.  The static
\vdnnAll and \vdnnConv policies are denoted as \texttt{all} and \texttt{conv}
in all the figures discussed in this section and are each evaluated with both
memory-optimal and performance-optimal (denoted as \texttt{(m)} and
\texttt{(p)}) convolutional algorithms across the network.  The baseline memory
manager (\texttt{base}) is similarly evaluated with both memory-optimal and
performance-optimal algorithms. The algorithms are dynamically chosen for
\vdnnDyn (denoted as \texttt{dyn}) as discussed in
\sect{sect:vdnn_transfer_policy}. Memory management policies that fail in
training the network, due to memory oversubscription, are marked with ($\ast$).

\subsection{GPU Memory Usage}
\label{sect:mem_usage}

Because \vdnn adopts a layer-wise memory allocation policy, the GPU memory
usage during forward/backward propagation will fluctuate depending on the
memory offloading policy chosen and the convolutional algorithm employed for a
given layer (\fig{fig:memory_usage_per_layer}).  We therefore discuss both the
maximum and average memory usage as shown in \fig{fig:vdnn_mem_usage_max_avg}.
The maximum memory usage corresponds to the largest memory allocated across the
entire run, which decides whether the target DNN application can be trained
at all.  The average memory on the other hand reflects how much memory
has been used on average, and conversely, freed up during forward/backward
propagation.  The smaller the average memory usage becomes, the more likely
\vdnn will have headroom to improve performance by: 1) employing
performance-efficient convolutional algorithms that require larger
workspace, and 2) reducing the total number of offload layers and prevent
potential performance drops due to offloading
(\fig{fig:vdnn_offload_prefetch_mechanism}).

\begin{figure}[t!]
\centering
\includegraphics[width=0.48\textwidth]{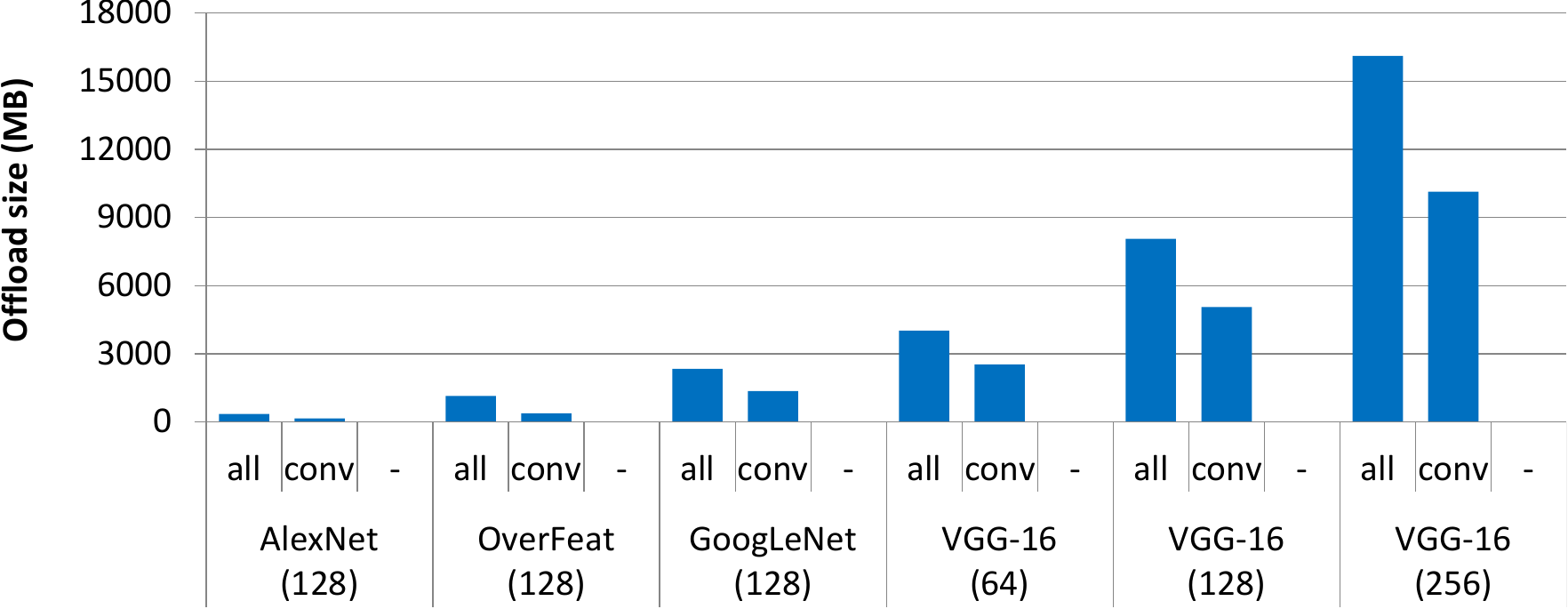}
\caption{The size of GPU memory allocations offloaded to host-side pinned memory using \texttt{cudaMallocHost()}.}
\vspace{-1em}
\label{fig:offload_traffic}
\end{figure}

Because the baseline policy provisions the memory
allocations to accommodate the entire network usage, the maximum and average
memory usage are identical. The baseline policy therefore is not able to train
networks like VGG-16 with batch 128 and 256 , which require more than the
physically available $12$ GB of memory.  Our \vdnn enhances the trainability of
a network by significantly reducing its memory requirements.  Overall, the
memory-optimal \vdnnAllM shows both the smallest average and maximum memory
usage as it always offloads a layer's input feature maps while using the most
memory-efficient algorithms.  As a result, \vdnnAll exhibits the highest
offload traffic being sent to host memory, reaching up to $16$ GB of GPU memory
savings for VGG-16~(256) (\fig{fig:offload_traffic}).  Such aggressive
offloading significantly improves memory efficiency and achieves an average
73\% and 93\% reduction on the maximum and average memory usage of the six networks
shown in \fig{fig:vdnn_mem_usage_max_avg}.  When employed with the
performance-optimal algorithm, the average memory savings of \vdnnAll are
slightly reduced to 64\% and 90\% for the maximum and average memory usage.
Because \vdnnConv only offloads the feature maps for the CONV layers, its
memory savings is not as high as \vdnnAll. However, \vdnnConv still reduces 
the maximum and average memory usage by 52\% and 76\% on average, even with the
performance-optimal algorithms employed across the network.

\vdnnDyn allocates the largest memory among the three \vdnn policies,
reducing the maximum and average memory consumption by 49\% and 69\% on average
compared to baseline. This is because \vdnnDyn tries to balance
its memory usage and performance, seeking to fit the network within GPU memory
while still optimizing performance by minimizing the number of offload
layers and employing the fastest possible convolutional algorithms.  The static
\vdnnAll and \vdnnConv, on the other hand, do not consider the overall
performance when the offloaded layers are chosen. For instance, VGG-16~(128)
trained with memory-optimal \vdnnAll only uses up to $4.8$ GB out of the $12$
GB of available memory. This configuration leads to a $61$\% performance loss
(\sect{sect:perf}) as \vdnnAll fails to exploit the remaining $7.2$ GB of the
memory for performance optimizations.  \vdnnDyn tries to bridge this gap by
dynamically deriving the offload layers as well as the best convolutional
algorithms to employ for each layer. We further discusses \vdnn's impact on performance in
detail at \sect{sect:perf}.

\subsection{Impact on Memory System}
\label{sect:dram_caching}

\begin{figure}[t!]
\centering
\includegraphics[width=0.48\textwidth]{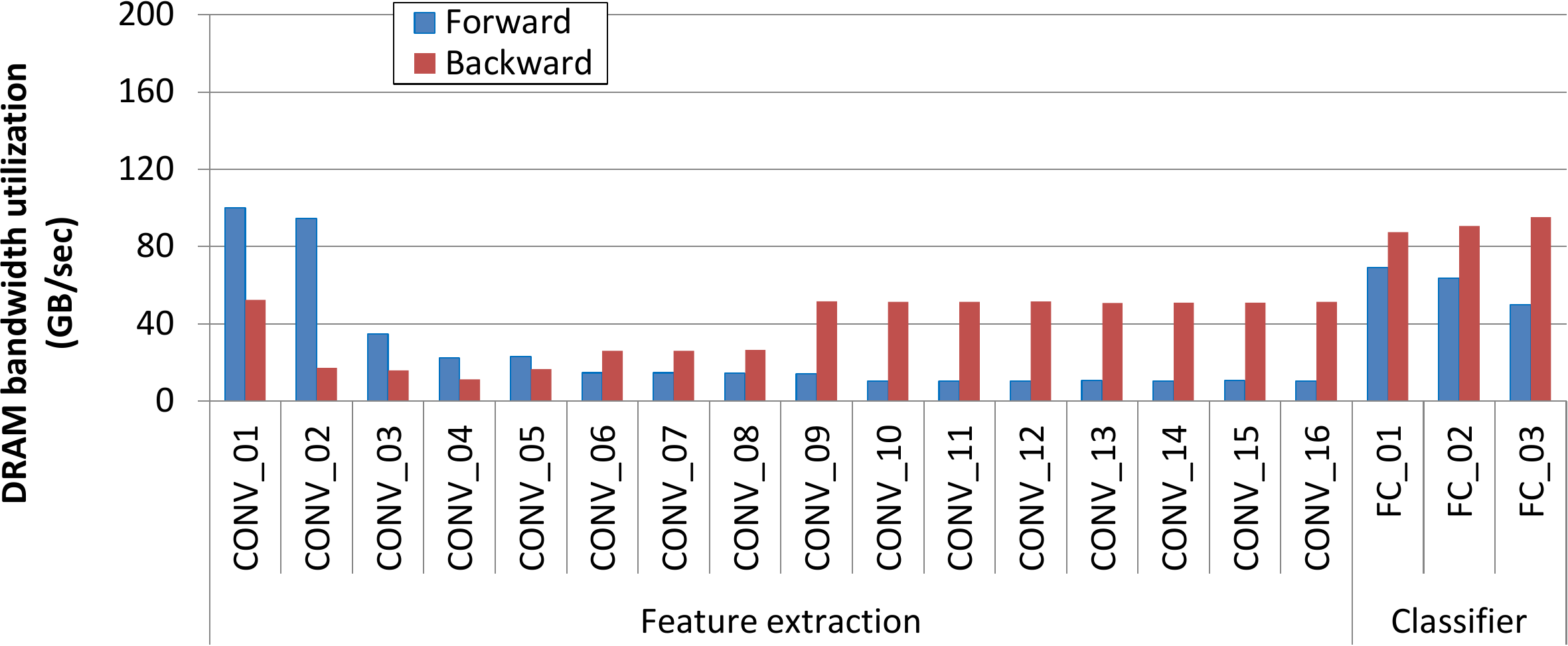}
\caption{Maximum DRAM bandwidth utilization for each CONV layer's forward and backward propagation.}
\vspace{-1em}
\label{fig:dram_bw}
\end{figure}

While \vdnn helps virtualize DNN's memory usage, it does come at the cost of
adding more read (offload) and write (prefetch) traffic to the GPU memory
subsystem, potentially interfering with the normal cuDNN operations.  Because
the additional \vdnn memory traffic can be up to the bandwidth of
the PCIe (maximum of $16$ GB/sec for gen3), its effect on 
performance will be determined by the normal cuDNN operation's memory bandwidth
intensity.  \fig{fig:dram_bw} shows the baseline's maximum DRAM bandwidth
utilization for VGG-16, which is measured separately for each CONV layer's
forward and backward propagation. The feature extraction layers rarely
saturate the $336$ GB/sec of peak memory bandwidth, providing more than enough
headroom for \vdnn's offload/prefetch traffic.  Even if a hypothetical, future
convolutional algorithm were to completely saturate the off-chip DRAM bandwidth, \vdnn's
additional traffic will approximately incur up to a worst-case (16/336) =
$4.7$\% performance overheads, which we believe is reasonable given the benefit
of virtualized memory.

\subsection{Performance}
\label{sect:perf}

\begin{figure*}[t!]
\centering
\includegraphics[width=0.995\textwidth]{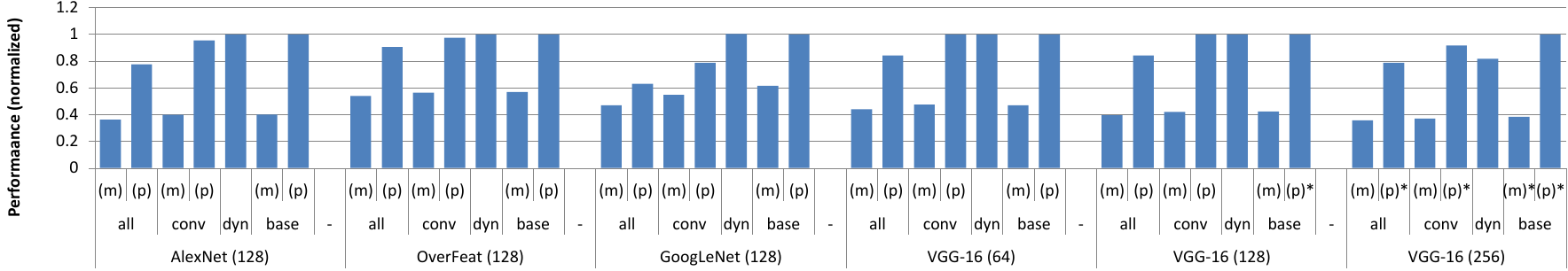}
\caption{Overall performance (normalized to baseline).}
\vspace{-1em}
\label{fig:vdnn_perf}
\end{figure*}

\fig{fig:vdnn_perf} summarizes the performance of \vdnn compared to baseline.
For a conservative evaluation, we only compare the latencies incurred in the
feature extraction layers because the classifier layers are executed
identically for baseline and \vdnn.  Because the baseline policy requires
more than $12$ GB of memory for VGG-16~(128) and VGG-16~(256) with
performance-optimal algorithms ($15$ GB and $28$ GB respectively), it is
impossible to train these two networks on a Titan X. We therefore establish an
\emph{oracular} baseline that removes the memory capacity bottlenecks of these
two networks for a conservative evaluation. The performance of this oracular
baseline is estimated by configuring all CONV layers with the fastest
algorithms and evaluating the latencies of each layers individually. The
latencies are later accumulated altogether to estimate overall performance.
Overall, \vdnnAll and \vdnnConv with memory-optimal algorithms exhibit an
average $58\%$ and $55\%$ performance loss (maximum $65\%$ and $63\%$
degradation) compared to baseline, an expected result as the memory manager puts
no effort into balancing memory usage and overall performance. The dynamic
\vdnnDyn does much better in terms of balancing memory efficiency and overall
throughput, closing the performance gap between the static \vdnn and baseline
and reaching an average 97\% of baseline's throughput (worst case 82\% of the oracular
baseline, for VGG-16~(256)).

\subsection{Power}
\label{sect:power}

This section discusses the effect of \vdnnDyn on overall GPU power consumption.
We use the system profiling utility of \texttt{nvprof}~\cite{cuda_profiler} to
measure the average and maximum GPU power consumption.  Each network is
executed for $50$ iterations of forward and backward propagation and the
reported average and maximum power consumption is averaged altogether.  All but
VGG-16~(128) have been executed with the performance-optimal convolutional
algorithms because VGG-16~(128) can only be trained with the memory-optimal
algorithms under baseline (\fig{fig:vdnn_mem_usage_max_avg}). Note that the
results for VGG-16~(256) are not discussed as this configuration can only be
trained with \vdnn, making it impossible to compare against baseline.  Overall,
\vdnnDyn incurs 1\% to 7\% maximum power overheads. As discussed in
\sect{sect:dram_caching}, the offload/prefetch memory traffic of \vdnn is one
of the biggest contributors to the instantaneous rise in peak power
consumption.  Nonetheless, the average power consumption (energy/time) is
rarely affected because of the following two factors: 1) \vdnnDyn does not
incur any noticeable performance overhead for these five networks, and
2) the studied DNNs rarely saturate the peak DRAM bandwidth
(\fig{fig:dram_bw}), so the additional energy overheads of \vdnn memory traffic
is expected to be negligible on average (\sect{sect:dram_caching}).

\begin{figure}[t!]
\centering
\includegraphics[width=0.48\textwidth]{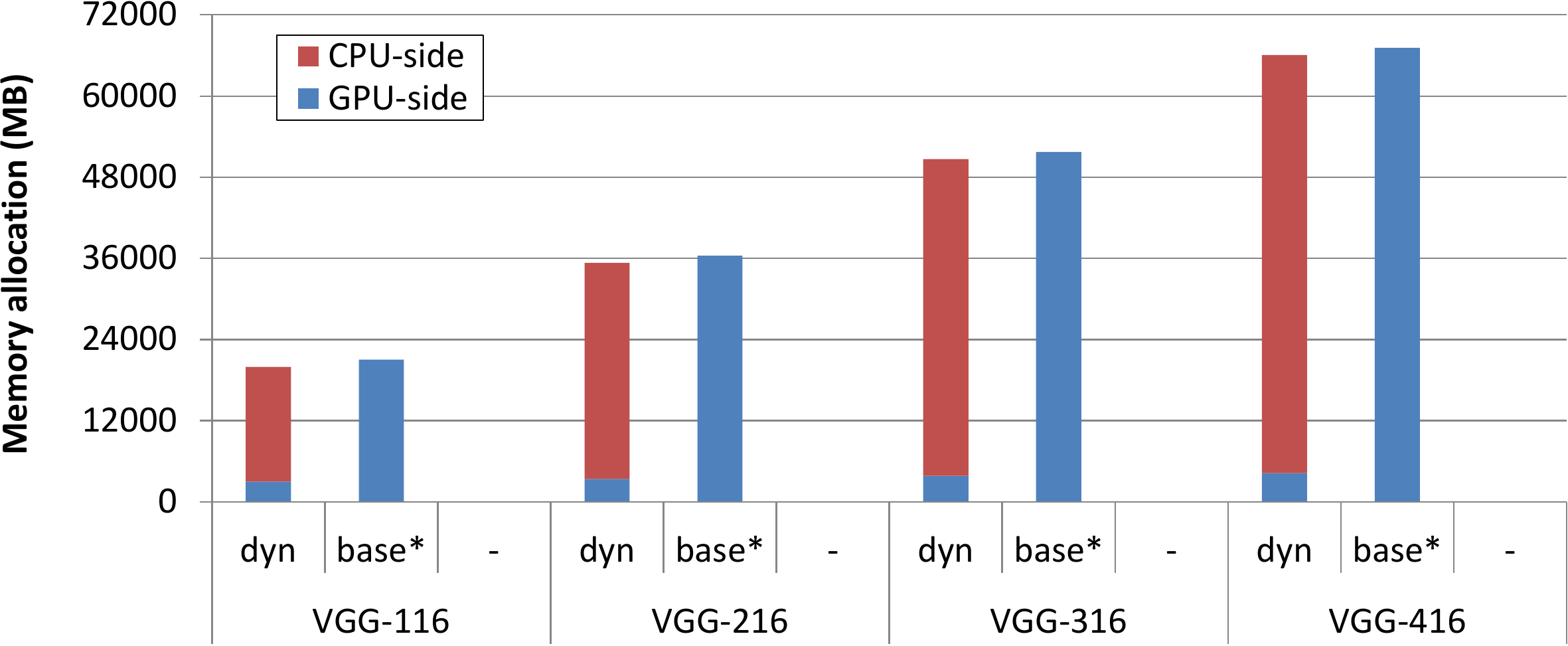}
\caption{CPU/GPU memory allocations for VGG-style, very deep networks (batch $32$) with \vdnnDyn and baseline.}
\vspace{-1em}
\label{fig:vdnn_case_study_deeper_vgg}
\end{figure}

\subsection{Case Study: Training Very Deep Networks}
\label{sect:trainability}

To highlight \vdnn's 
scalability in training very deep networks, we perform a case study on four
VGG-style networks that contain hundreds of CONV layers and scale up the
network memory requirements. As mentioned in \sect{sect:benchmark}, the batch
size is set to be much smaller than those studied in previous subsections
(which ranges from batch $128$ to $256$) as means to highlight the memory
scaling effect of layer depth despite its small batch size.
\fig{fig:vdnn_case_study_deeper_vgg} shows the memory allocation requirements
of baseline and \vdnnDyn for these very deep neural networks.  As the number of
CONV layers increases from $16$ to $416$, the baseline memory requirements
monotonically increase by $14$ times (from $4.9$ GB to $67.1$ GB), even with a
small batch size of $32$.  Thanks to its layer-wise memory allocation policy,
\vdnnDyn significantly reduces the memory usage of all four networks, only
using up to $4.2$ GB of GPU memory and having all remaining $81$\% to $92$\% of
overall memory allocations to reside in CPU memory. Compared to the oracular
baseline, \vdnnDyn also did not incur any noticeable performance degradations
because the offload and prefetch latency is completely hidden inside the layer's
DNN computations while still being able to employ the performance-optimal
algorithms across the network.

%% file: tex/related.tex
\section{Related Work}
\label{sect:related_work}

There have been a variety of proposals aiming to reduce the memory usage of
neural networks. Network pruning
techniques~\cite{hanson:1989:mnc,lecun:1990:brain_damage,hassibi:1993:brain_surgeon,song:2015:pruning,song:2015:compression}
remove small valued weight connections from the network as means to reduce
network redundancy, leading to a reduction in memory consumption.  Another
redundancy mitigating approach uses quantization~\cite{vanhoucke:2011:snn} or
reduced precision~\cite{judd:2016:reducedPrecision} to reduce the number of
bits required to model the network.  A variety of network compression
techniques have also been explored by Gong et al.~\cite{gong:2014:compression}
to reduce the memory usage of DNNs.

Although these prior studies reduce DNN memory requirements, they fall short in
several respects. First, weights only account for a small fraction of the
memory usage in state-of-the-art DNNs, as shown in
\fig{fig:mem_usage_categorization}. Thus, proposals that optimize memory usage
of weights, while beneficial in terms of memory bandwidth utilization and
energy-efficiency, provide only limited opportunity for memory capacity
savings. Second, using reduced precision occasionally results in loss of
classification accuracy unless carefully tuned for the given network and task.
Our proposal optimizes the memory consumptions of the intermediate feature maps
which are the most dominant data structures in DNNs.

Several prior works discussed mechanisms to support virtualized memory on GPUs.
Pichai et al.~\cite{gpu_tlb} and Power et al.~\cite{gpu_x86_at} proposed TLB
implementations that consider the unique memory access patterns of
GPUs, improving the throughput of address translations as well as overall
system throughput.  Zheng et al.~\cite{gpu_paging} discuss features needed 
in the GPU hardware and software stack to close the performance gap of GPU paged memory
versus legacy programmer-directed memory management techniques.
As discussed in \sect{sect:motivation}, page-migration based virtualization solutions
are likely to underutilize PCIe bandwidth significantly and incur performance overheads 
when training networks that oversubscribe GPU memory.

While less
directly related to \vdnn, a variety of accelerator architectures have also
been proposed for DNNs~\cite{diannao,dadiannao,eyeriss,song:2015:eie,eyeriss_isca,redeye,minerva,dnn_pim_reram,isacc,cnvlutin,shidiannao}.
While these custom ASIC designs drastically improve the energy-efficiency of
DNNs, none of these address the memory capacity bottlenecks of DNN
training, a unique contribution of our work.

%% file: tex/conclusion.tex
\section{Conclusion}

Existing machine learning frameworks require users to carefully manage their
GPU memory usage so that the network-wide memory requirements fit
within the physical GPU memory size. We propose \vdnn, a scalable, memory-efficient runtime memory manager that
virtualizes the memory usage of a network across CPU and GPU memories. Our
\vdnn solution reduces the average GPU memory usage of AlexNet by up to 89\%,
OverFeat by 91\%, and GoogLeNet by 95\%, substantially improving the memory-efficiency of DNNs.
Similar experiments to VGG-16~(256) result in an average 90\% reduction in memory
usage at a cost of 18\% performance penalty compared to an oracular baseline. 
We also study the scalability of \vdnn to extremely deep neural networks, 
showing that \vdnn can train networks
with hundreds of layers without any performance loss.